\documentclass{article}

\usepackage{arxiv}

\usepackage[utf8]{inputenc} % allow utf-8 input
\usepackage[T1]{fontenc}    % use 8-bit T1 fonts
\usepackage{hyperref}       % hyperlinks
\usepackage{url}            % simple URL typesetting
\usepackage{booktabs}       % professional-quality tables
\usepackage{amsfonts}       % blackboard math symbols
\usepackage{nicefrac}       % compact symbols for 1/2, etc.
\usepackage{microtype}      % microtypography
\usepackage{lipsum}		% Can be removed after putting your text content
\usepackage{graphicx}
\usepackage{natbib}
\usepackage{doi}

\usepackage{graphicx}
\usepackage{graphicx}
\usepackage{algorithm} 
\usepackage{algorithmic}
\usepackage{multirow,lscape,rotating,url,amsfonts,amsmath,empheq,fancybox}
\usepackage{pstricks,pst-node,pst-plot,pst-coil}
\usepackage{changes}
\usepackage{caption}
\usepackage{tabularx}
\usepackage{xltabular}
\usepackage[short]{optidef}
\usepackage{amssymb}
\usepackage{mathtools}
\usepackage{blkarray, bigstrut} %
\usepackage{booktabs}

\usepackage{mathptmx}
\usepackage{longtable}
\usepackage{listings}
\usepackage{subcaption}
\usepackage{bbm}
\usepackage{xcolor,colortbl}
\usepackage{url}
\usepackage{soul}\setuldepth{article}
\usepackage{amsmath}
\usepackage{blkarray}%
\usepackage{csquotes}
\allowdisplaybreaks

\newcommand{\VNSfull}{VNS$_{\mathrm{full}}$}
\newcommand{\VNSred}{VNS$_{\mathrm{red}}$}
\title{Mathematical Formulations And Results Regarding Two Echelon Electric Vehicle Routing Problems}

\author{ 
    \href{https://orcid.org/0000-0001-7376-7008}{\includegraphics[scale=0.06]{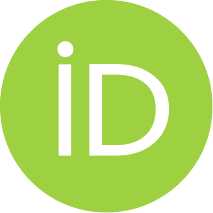}\hspace{1mm}Mehmet Anıl Akbay}\thanks{Corresponding author.} \\
    Artificial Intelligence Research Institute (IIIA-CSIC) \\
    Campus UAB, Bellaterra, Spain \\
    \texttt{makbay@iiia.csic.es} \\
    \And
    \href{https://orcid.org/0000-0002-1736-3559}{\includegraphics[scale=0.06]{orcid.pdf}\hspace{1mm}Christian Blum} \\
    Artificial Intelligence Research Institute (IIIA-CSIC) \\
    Campus UAB, Bellaterra, Spain \\
    \texttt{christian.blum@csic.es} \\
}

\renewcommand{\shorttitle}{\textit{arXiv} Template}

\hypersetup{
pdftitle={A template for the arxiv style},
pdfsubject={q-bio.NC, q-bio.QM},
pdfauthor={David S.~Hippocampus, Elias D.~Striatum},
pdfkeywords={First keyword, Second keyword, More},
}

\begin{document}
\maketitle

\begin{abstract}
The growing need for sustainable logistics solutions has led to the evolution of vehicle routing problems (VRPs) into more complex variants that address modern challenges. Among these, the Two-Echelon Electric Vehicle Routing Problem (2E-EVRP) has emerged as a significant problem variant, integrating electric vehicles (EVs) into a multi-echelon distribution system. This problem considers environmental and operational constraints such as limited battery range, charging infrastructure, and urban logistics complexities. In this report, we present a comprehensive mathematical formulation for the 2E-EVRP and its variants, which include constraints like time windows, simultaneous pickup and delivery, and partial deliveries. These formulations aim to provide a robust framework for optimizing multi-tiered distribution networks using sustainable practices. Computational experiments demonstrate the effectiveness of the proposed methods. 

\end{abstract}

% keywords can be removed

\section{Introduction}

The increasing demand for efficient logistics and transportation solutions, coupled with the need for environmentally sustainable practices, has driven the evolution of traditional vehicle routing problems (VRPs) into more complex and specialized variants. Introduced by~\cite{dantzig1959truck}, the classic VRP aims to optimize the delivery of goods from a central depot to a set of customers while minimizing travel costs. The VRP can be viewed as an extension of the Traveling Salesman Problem (TSP), which seeks the shortest path for a single vehicle to visit all customers exactly once. In contrast, the VRP involves multiple vehicles, each subject to capacity constraints, adding considerable complexity to the problem. Since its inception, numerous variants of VRPs have been developed to accommodate a wide range of real-world logistics problems.

\begin{figure}[ht] \centering \includegraphics[width=1.0\textwidth]{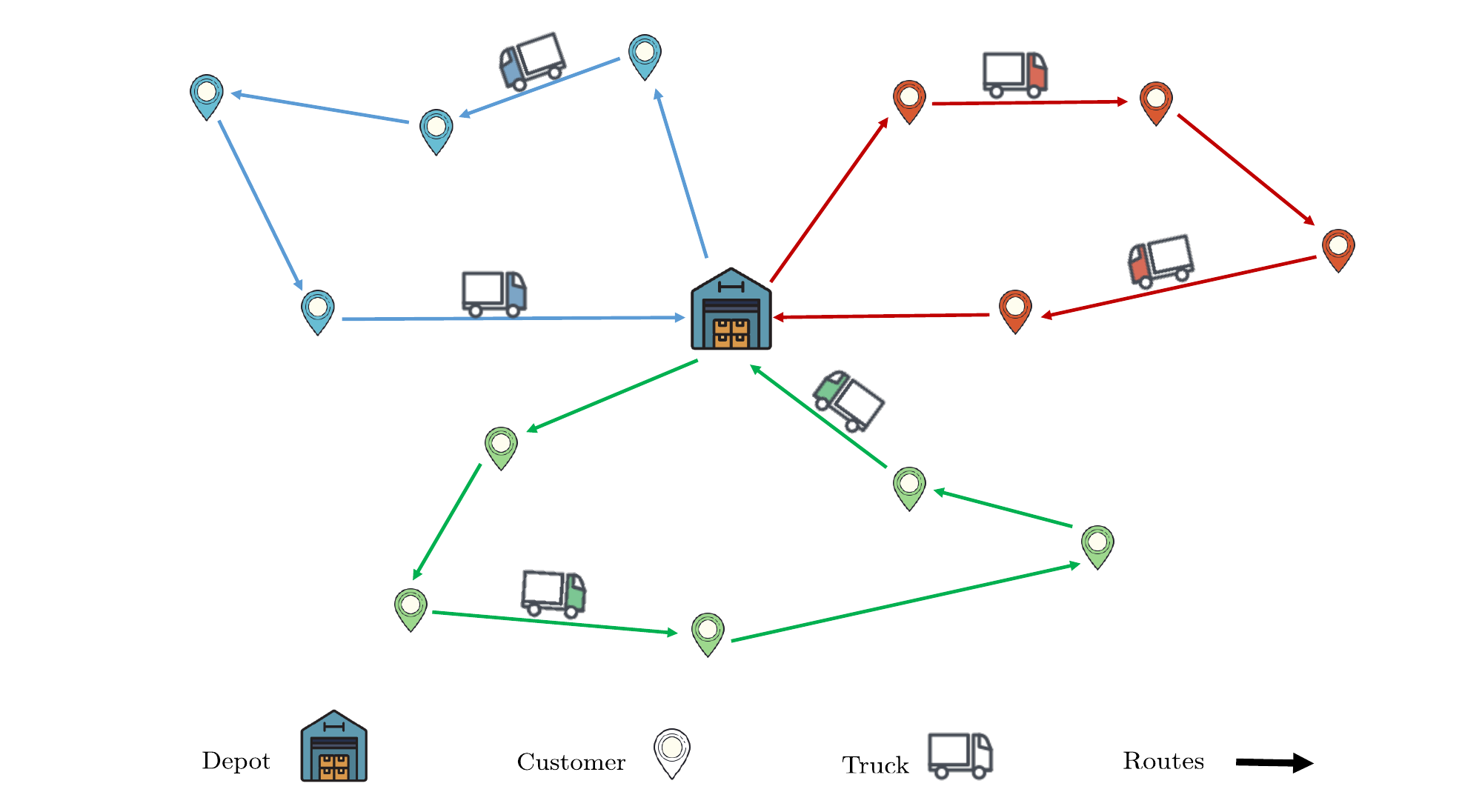} \caption{Illustration of a Capacitated Vehicle Routing Problem. Icons by Icons8 (icons8.com).} \label{fig:intro1}
\end{figure}

Among these variants, the Electric Vehicle Routing Problem (EVRP) and the Two-Echelon Vehicle Routing Problem (2E-VRP) have gained particular attention due to their relevance in modern sustainable logistics practices. The EVRP, as first formulated by~\cite{conrad2011recharging}, integrates electric vehicles (EVs) into the classic VRP framework. EVs are increasingly being adopted in the logistics industry due to their potential to reduce greenhouse gas emissions and dependence on fossil fuels~\citep{Lellis2021}. However, the limited driving range and the need for recharging introduce additional constraints, requiring strategic planning for battery management. The objective of the EVRP is not only to minimize traditional costs, such as travel distance and time but also to account for the unique characteristics of EVs, such as energy consumption and recharging schedules~\citep{erdougan2012green, schneider2014electric}.

\begin{figure}[ht] \centering \includegraphics[width=1.0\textwidth]{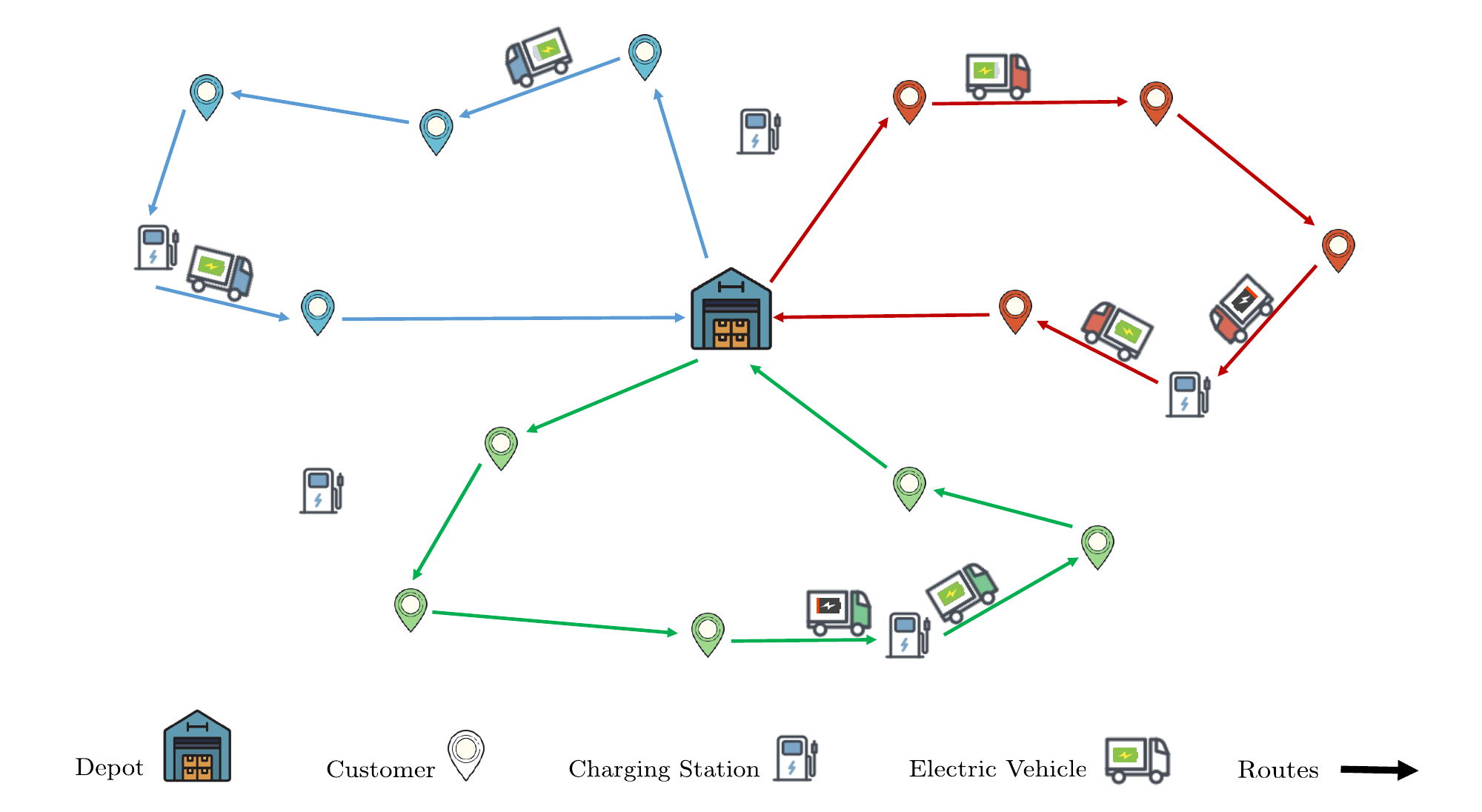} \caption{Illustration of an Electric Vehicle Routing Problem. Icons by Icons8 (icons8.com).} \label{fig:intro2}
\end{figure}

On the other hand, the Two-Echelon Vehicle Routing Problem (2E-VRP) addresses the complexities of multi-tiered distribution systems, which are particularly relevant in urban logistics. The 2E-VRP divides the delivery process into two distinct stages or echelons. In the first echelon, goods are transported from a central warehouse to intermediate facilities called satellites, typically located on the outskirts of urban areas. The second echelon involves the delivery of goods from these satellites to the final customers using smaller vehicles suited for navigating city environments. The 2E-VRP aims to optimize transportation costs across both echelons while considering the constraints imposed by urban logistics, such as vehicle access restrictions and limited road widths~\citep{crainic2004advanced, perboli2011two}.

\begin{figure}[ht] \centering \includegraphics[width=1.0\textwidth]{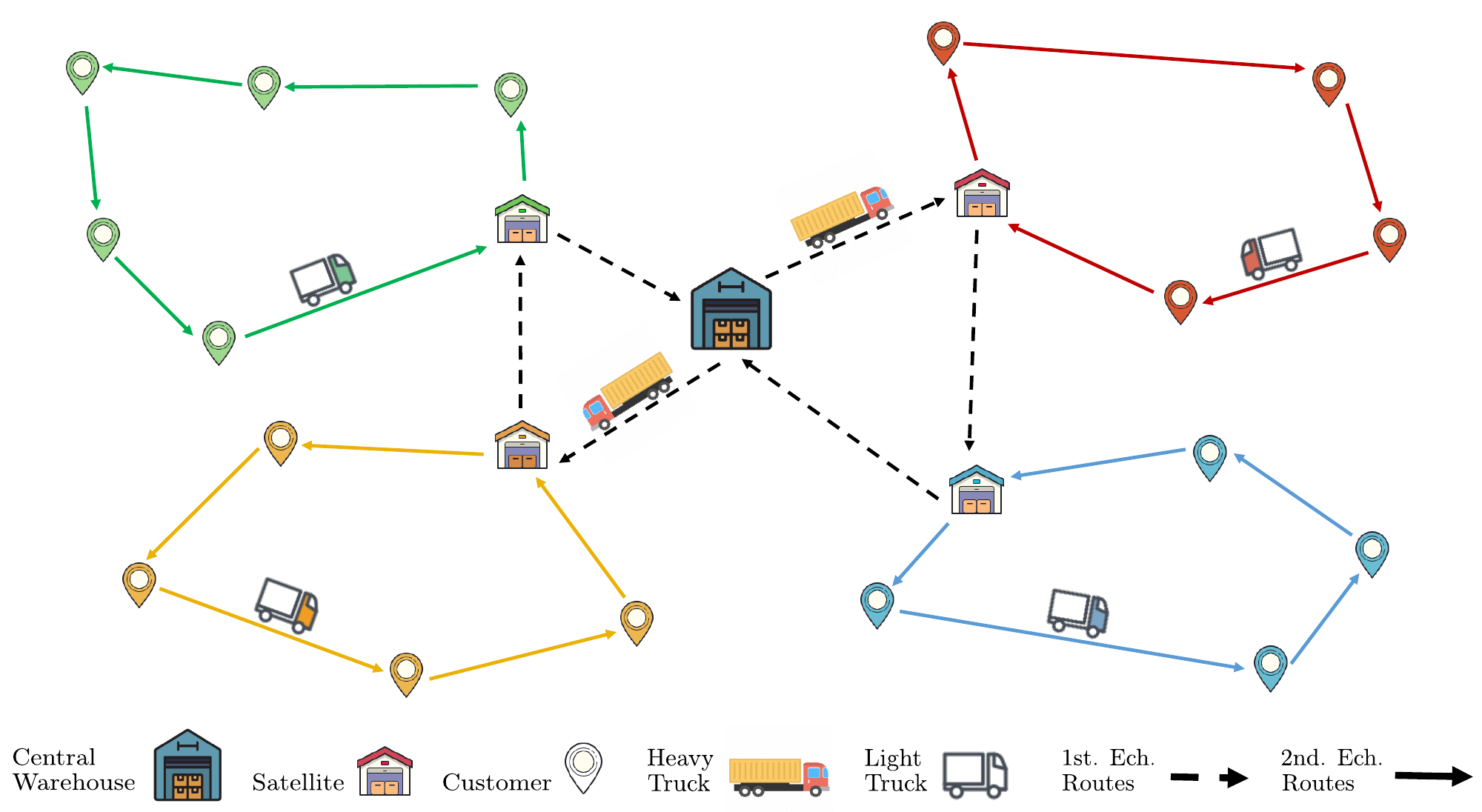} \caption{Illustration of a Two-Echelon Vehicle Routing Problem. Icons by Icons8 (icons8.com).} \label{fig:2E-VRP_Network} \end{figure}

Building on these two research lines, the Two-Echelon Electric Vehicle Routing Problem (2E-EVRP) integrates the use of electric vehicles into a multi-echelon distribution framework. The 2E-EVRP aims to leverage the environmental benefits of EVs while addressing the complexities of multi-tiered delivery systems. In the first echelon, conventional trucks transport goods from a central warehouse to satellites, where the goods are then redistributed by EVs in the second echelon to the final customers within urban areas. Integrating EVs into the two-echelon structure aligns with the growing emphasis on sustainable urban logistics, reducing emissions and noise pollution in city centers~\citep{breunig2019electric}.

\begin{figure}[ht] \centering \includegraphics[width=1.0\textwidth]{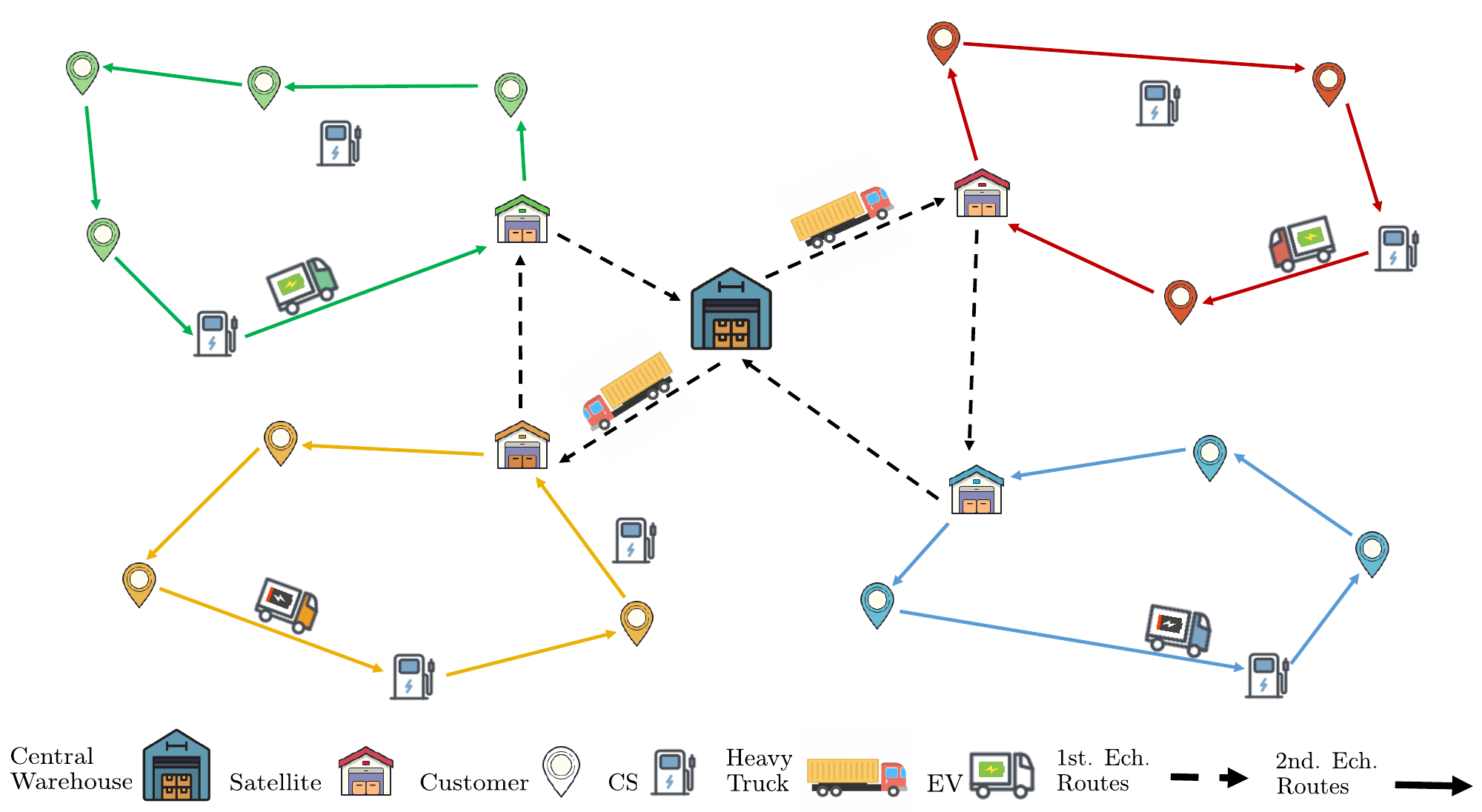} \caption{Illustration of a Two-Echelon Electric Vehicle Routing Problem. Icons by Icons8 (icons8.com).} \label{fig:2E-EVRP} \end{figure}

The 2E-EVRP introduces several unique challenges, combining aspects of the EVRP and the 2E-VRP. In the first echelon, the key challenge lies in the efficient allocation of customers to satellites and the coordination of deliveries from the warehouse. The second echelon involves additional complexities related to the limited range of EVs, the availability of charging infrastructure, and operational considerations such as time windows for customer deliveries. The primary objective of the 2E-EVRP is to minimize the total cost of the distribution process, encompassing travel distances, vehicle costs, operational expenses, and energy consumption.

The literature on 2E-EVRPs is still relatively limited compared to other VRP variants. Early studies by~\cite{jie2019two} introduced a 2E-EVRP with battery-swapping stations, proposing a hybrid algorithm combining column generation and large neighborhood search (LNS). \cite{breunig2019electric} developed a metaheuristic based on LNS and an exact decomposition approach for solving the 2E-EVRP. Other researchers have explored various extensions of the problem, such as incorporating time windows and partial recharging of EVs~\citep{cao2021heterogeneous, wu2023branch}. Recent developments include the work of ~\cite{wang2021two}, who introduced a MILP model for the 2E-EVRP with battery-swapping stations, highlighting the importance of integrating charging infrastructure planning into the routing strategy.

In this report, we present a comprehensive study of the 2E-EVRP and its variants, focusing on real-world logistics constraints such as time windows, simultaneous pickup and deliveries (SPD), and partial deliveries. We start by defining the baseline problem, the 2E-EVRP, and providing a detailed mathematical formulation. We then extend this formulation to address additional complexities introduced by the variants, highlighting the modifications needed to incorporate these features. The goal is to provide a robust framework for modeling and solving 2E-EVRPs, contributing to the ongoing development efforts.

\section{Data Description}
\label{sec:data_description}

The set of problem instances used in this study extends the classical EVRP benchmark instances introduced by Schneider et al.~\cite{schneider2014electric}, originally derived from Solomon's VRP-TW instances~\cite{solomon1987algorithms}. These instances have been modified to reflect the characteristics of the 2E-EVRP and its variants, incorporating additional real-world constraints such as time windows, simultaneous pickup and delivery (SPD), and partial deliveries.

The dataset comprises three spatial distribution types: Clustered (C), where customers are arranged in dense clusters; Random (R), where customers are randomly distributed across the region; and Random-Clustered (RC), which combines clustered and randomly placed customers. Each instance includes nodes representing central warehouses (CW), the starting points for first-echelon deliveries; satellites, which are intermediate facilities where goods are transferred to electric vehicles; customers, who are the endpoints for deliveries and may have demands involving SPD or partial deliveries; and charging stations, where electric vehicles can recharge.

The dataset used in this study contains small instances with 5, 10 customers and 1 satellite, and 15 customers and 2 satellites. Large instances include 100 customers and 8 satellites.

The dataset used in this study is publicly available~\cite{mehmet_anil_akbay_2024_14215974}, and a detailed description of the methodology for its modification can be found in~\citep{akbay2024benchmark}.

\section{Problem Description}
\label{sec:problem}

In the following, we provide a technical description with a MILP model of the 2E-EVRP and its variants considering various real-world logistics constraints. We will first present the formulation of the baseline problem, 2E-EVRP, and then we will present the other variants by highlighting the modifications needed to be done on the basic model by referring to the basic model. For this purpose, we first define the sets and notations required in the models and shown in Table~\ref{tab:sets}.

\begin{table}[!t]
    \centering
    \caption{Sets and notations}
    \label{tab:sets}
\scalebox{0.9}{
\begin{tabular}{ll} \hline
    $n_d$, $n_s$ , $n_{r}$ , $n_c$ &: Number of central warehouses, satellites, charging stations and customers, respectively \\
    $N_D$   &: Set of central warehouses, $N_D = \{ n_{d_1},...,n_{d_{n_d}} \}$ \\
    $N^{'}_{D}$  &: Set of dummy central warehouses corresponding to $N_D$, $N^{'}_{D} = \{ n^{'}_{d_1},...,n^{'}_{d_{n_d}} \}$\\
    $N_{S}$   &: Set of satellites, $N_S = \{ n_{s_1},...,n_{s_{n_s}} \}$  \\
    $N^{'}_{S}$  &: Set of dummy satellites corresponding to $N_S$, $N^{'}_{S} = \{ n^{'}_{s_1},...,s^{'}_{s_{n_s}} \}$ \\
    $N_R$     &: Set of charging stations, $N_S = \{ n_{r_1},...,n_{r_{n_r}} \}$  \\
    $N_C$   &: Set of customers, $N_S = \{ n_{c_1},...,n_{c_{n_c}}\}$ \\
    $N_{DS}$   &: Set of central warehouses and satellites, $N_{DS} = N_D \cup N_S$  \\
    $N_{SD}$   &: Set of satellites and dummy central warehouses, $N_{SD} = N_S \cup N^{'}_{D}$ \\
    $N_{DSD}$   &:  Complete set of nodes in the first echelon, $N_{DSD} = N_D \cup N_S \cup N^{'}_{D}$ \\
    $N_{RC}$  &: Set of charging stations and customers, $N_{RC} = N_R \cup N_{C}$  \\
    $N_{SRC}$     &: Set of satellites, charging stations and customers, $N_{SRC} = N_S \cup N_R \cup N_C$  \\
    $N_{RCS}$     &: Set charging stations, customers and dummy satellites, $N_{SRC} = N_R \cup N_C \cup N^{'}_{S}$ \\
    $N_{SRCS}$     &: Complete set of nodes in the second echelon, $N_{SRCS} = N_{S} \cup N_R \cup N_C \cup N^{'}_{S}$ \\
    \hline
\end{tabular}}
\end{table}

\subsection{Baseline Problem: Two Echelon Electric Vehicle Routing Problem}
\label{sec:base-problem}

2E-EVRPs can be defined on a complete, directed graph $G(N,A)$ that is formed by the following subsets of nodes: the set of central warehouses (also called depots) $(N_D)$, the set of satellites $(N_S)$, the set of charging stations $(N_R)$, and the set of customers $(N_C)$. Note that $N_S$ and $N_R$ also include multiple copies of each satellite and charging station to allow multiple visits to any of the satellites and charging stations. The set of arcs on the other hand ($A$) includes (1) arcs that connect central warehouses and satellites \mbox{$A^{1}=\{(i,j) \mid i \neq j \text{ and } i,j \in N_{DSD}\}$} and (2)~arcs that connect satellites, customers and charging stations $A^{2}=\{(l,m) \mid l \neq m$ $\text{ and } l,m \in N_{SRCS} \}$. Each arc $(i,j) \in A^{1}$ is associated with a distance $d^{1}_{ij}$ and each arc $(l,m) \in A^{2}$ is associated with a distance $d^{2}_{lm}$.

Two different fleets of vehicles, each one homogeneous in itself, serve in the first and second echelons in order to meet customer demands. A fleet of large trucks with internal combustion engines are located in a central warehouse and transfer products between the central warehouses and the satellites, while a fleet of electric vehicles is present at the satellites and transfer products between satellites and customers (demand points). In the first echelon, a truck with a loading capacity of $Q^{1}$ starts its tour from a central warehouse, visits one or more satellites, and returns to the central warehouse from which the tour started. Not all satellites have to be visited by large trucks unless there is a customer demand. Furthermore, a satellite can be visited by multiple large vehicles if the demand of the satellite exceeds the vehicle capacity. In the second echelon, on the other hand, each customer with a delivery demand $D^{2}_i > 0$ must be served by an electric vehicle with a loading capacity of $Q^{2}$. An electric vehicle starts its tour with a fully charged battery ($B$) and the vehicle's battery is consumed in proportion to the distance travelled. The constant $h$ represents the battery consumption rate of an electric vehicle per unit distance travelled. If a charging station is visited, the electric vehicle's battery is fully charged up to level $B$ with a constant charging rate of $g > 0$.

Our MILP model contains the following binary decision variables. A decision variable $x_{ij}$ takes value 1 if arc $(i,j) \in A^{1}$ is traversed, and 0 otherwise. Moreover, a decision variable $y_{ij}$ takes value 1 if arc $(i,j) \in A^{2}$ is traversed, and 0 otherwise. Next, decision variables $BSCa_i$ and $BSCd_i$ record the battery state of charge on arrival, respectively departure, at (from) vertex $i\in N_{SRCS}$. Furthermore, for each arc $(i,j) \in A^{1}$, variable $u^{1}_{ij}$ denotes the remaining cargo to be delivered to satellites of the route. Similarly, for each arc $(i,j) \in A^{2}$, variable $u^{2}_{ij}$ denotes the remaining cargo for the route. Since the demand of each satellite depends on the customers serviced through it, decision variables $D^{1}_i$ is introduced to calculate the delivery demands of satellites. Finally, variable $z_{ij}$ takes value 1 if customer $(i)$ is serviced from satellite $(j)$, and 0 otherwise. The MILP model can then be stated as follows.

%\pagebreak 
{\small 
    \begin{align}   
    \textbf{Min}    \quad  & \sum\limits_{i \in N_{DS}} \sum\limits_{j \in N_{DSD}} { d^1_{ij}*x_{ij}} + \sum\limits_{l \in N_{SRC}} \sum\limits_{m \in N_{SRCS}} { d^2_{lm}*y_{lm}} +\sum\limits_{j \in N_{DSD} } {x_{0j}}*c^{lv} +  \sum\limits_{i \in N_{S}}\sum\limits_{j \in N_{SRCS} } {y_{ij}}*c^{ev}  \label{m1:1} 
    \end{align}\vspace{-15pt}
    }%
    {\small  
    \begin{align}
     %\textbf{s.t.} 
        \quad  & \sum\limits_{j \in N_{SD}}{x_{ij} \leq 1} &  \forall i \in N_S  \label{m1:2}    \\
        \quad  & \sum\limits_{i \in N_{DS}, i \neq j}{x_{ij}}-\sum\limits_{i \in N_{SD}, i \neq j}{x_{ji}} =  0  &  \forall j \in N_S   \label{m1:3}    \\
        \quad  & \sum\limits_{i \in N_{DS}, i \neq j}{u^{1}_{ij}}-\sum\limits_{i \in N_{SD}, i \neq j}{u^{1}_{ji}} =  D^{1}_{j}  &  \forall j \in N_S   \label{m1:4}    \\
        \quad  & 0 \leq u^{1}_{ij} \leq Q^{1}  &  \forall i \in N_D, j\in N_{DS}   \label{m1:5}  \\   
        \quad  & u^{1}_{ij} \leq Q^{1}*x_{ij}  & \forall i \in N_D, j\in N_{DS}, i \neq j \label{m1:6}    \\
        \quad  & \sum\limits_{l \in N_{C}}{z_{li} * D^{2}_l} = D^{1}_{i}  &  \forall i \in N_S   \label{m1:7} \\
        \quad  & \sum\limits_{j \in N_{RCS}, i \neq j}{y_{ij}} = 1  &  \forall i \in N_C   \label{m1:8} \\
        \quad  & \sum\limits_{j \in N_{RCS}, i \neq j}{y_{ij}} \leq 1  &  \forall i \in N_R   \label{m1:9} \\
        \quad  & \sum\limits_{i \in N_{SRC}, i \neq j}{y_{ij}}-\sum\limits_{i \in N_{RCS}, i \neq j}{y_{ji}} =  0  &  \forall j \in N_{RC}   \label{m1:10}    \\
        \quad  & \sum\limits_{i \in N_{S}}{z_{li} = 1} &  \forall l \in N_{RC}  \label{m1:11} \\
        \quad  & y_{li} \leq z_{li}  &  \forall i \in N_S, l \in N_{RC}   \label{m1:12}  \\    
        \quad  & y_{il} \leq z_{li}  &  \forall i \in N_S, l \in N_{RC}   \label{m1:13}  \\    
        \quad  & y_{lm}+z_{li}+\sum\limits_{s \in N_{S}, i\neq s}{z_{ms}} \leq 2 &  \forall l,m \in N_{RC}, l \neq m, \forall i \in N_S   \label{m1:14} \\
        \quad  & \sum\limits_{i \in N_{SRC}, i \neq j}{u^{2}_{ij}}-\sum\limits_{i \in N_{RCS}, i \neq j}{u^{2}_{ji}} =  D^{2}_{j}  &  \forall j \in N_{RC}   \label{m1:15}    \\
        \quad  & 0 \leq u^{2}_{ij} \leq Q^{2}  &  \forall i \in N_S, j\in N_{RCS}   \label{m1:16}  \\   
        \quad  & u^{2}_{ij} \leq Q^{2}*y_{ij}  &  \forall i \in N_{RCS}, j\in N_{SRC}   \label{m1:17}  \\   
        \quad  & 0 \leq BSCa_{j} \leq BSCa_{i}-(hd_{ij})y_{ij}+B(1-y_{ij}) &  \forall i \in N_C, \forall j \in N_{RCS}, i\neq j   \label{m1:18}    \\
        \quad  & 0 \leq BSCa_{j} \leq BSCd_{i}-(hd_{ij})y_{ij}+B(1-y_{ij}) &  \forall i \in N_{SR}, \forall j \in N_{RCS}, i\neq j   \label{m1:19}    \\
        \quad  & BSCa_{i} \leq BSCd_{i} \leq B & \forall i \in N_{SR}  \label{m1:20}  \\
        \quad  & x_{ij} \in {0,1} &  \forall i \in N_{SRC}, j \in N_{RCS}, l\neq m  \label{m1:21} \\
        \quad  & y_{lm} \in {0,1}  &  \forall l \in N_{SRC}, m \in N_{RCS}, l\neq m  \label{m1:22}
    \end{align}
}%

In this problem, solutions using fewer vehicles---that is, with fewer routes---are preferred over others, even if the total distance traveled is higher than in other routes. Therefore, the objective function does not only consider the traveled distance but also adds an extra cost $c^{lv}$ for each large vehicle used in the first echelon and $c^{ev}$ for each electric vehicle used in the second echelon. Note, in this context, that the number of large vehicles used in a solution is equal to the number of variables on outgoing arcs of a central warehouse with a value of $1$. Moreover, the number of electric vehicles used in a solution is equal to the number of variables on outgoing arcs of satellites that have a value of $1$. In this way, the objective function~(\ref{m1:1}) minimizes the sum of the total distance traveled and the vehicle costs. Constraints~(\ref{m1:2}) control the connectivity of satellites, and constraints~(\ref{m1:3}) guarantee the balance of flow in the first echelon nodes. Constraints~(\ref{m1:4}) guarantee that the delivery demands of satellites are satisfied by the large vehicles serving in the first echelon. Constraints~(\ref{m1:7}) determine each satellite’s delivery demand to be the total delivery demands of those customers served by the relevant satellite. Constraints~(\ref{m1:8}) and~(\ref{m1:9}) control the connectivity of customers and charging stations. Constraints~(\ref{m1:10}) ensure the balance of flow for the second echelon nodes. Constraints~(\ref{m1:11}) guarantee that a customer receives service from only one satellite. Constraints~(\ref{m1:12})-(\ref{m1:14}) ensure that a tour started from a satellite ends at the same satellite. Constraints~(\ref{m1:15})-(\ref{m1:17}) guarantee that the delivery demands of customers are satisfied by the electric vehicles serving in the second echelon. Finally, constraints~(\ref{m1:18})-\ref{m1:20}) are battery state constraints.

\subsubsection{Computational Results}

In the following, we provide a detailed comparison of selected methods for solving the above base problem. First, we applied the integer linear programming (ILP) solver CPLEX (version 12.10) to all problem instances. CPLEX was given a time limit of 2 hours of CPU time for each problem instance. Additionally, we applied the CMSA (Construct, Merge, Solve, and Adapt) variant and the probabilistic implementations of two constructive heuristics (C\&W Savings and Insertion heuristics) from~\cite{akbay2023application} by reducing them from the more complex Two-Echelon Electric
Vehicle Routing Problem with Simultaneous Pickup and Deliveries (2E-EVRP-SPD) to the 2E-EVRP. The problem instances utilized are those outlined in Section~\ref{sec:data_description}. Note that the same parameter values as reported in~\cite{akbay2020parallel} were used for this purpose. CMSA was run with a computation time limit of 150 CPU seconds for small problem instances and 900 CPU seconds for large problem instances. Moreover, CMSA was executed 10 times on each problem instance to account for its stochastic nature.

Tables~\ref{base_tab:results1}--\ref{base_tab:results3} show the numerical results for small problem instances with 5, 10, and 15~customers, respectively. The structure of these tables is as follows. Instance names are given in the first column. The subsequent columns present the results for the three approaches under consideration: CPLEX, the probabilistic constructive heuristics, and CMSA.

For each approach, the 'Dist' column provides the objective function value of the solutions generated. For CMSA and the probabilistic constructive heuristics, this represents the best solution found across 10 independent runs. Additionally, a column with the heading 'Avg' provides the average objective function value across these 10 runs. The columns with the heading '$\overline{t(s)}$' indicate the computation time taken by each method to find its best solution. Finally, the 'Gap(\%)' column provides the gap (in percent) between the best solution and the best lower bound found by CPLEX. A gap value of zero indicates that CPLEX found an optimal solution.

When considering the smaller-sized 2E-EVRP instances, CPLEX managed to provide optimal solutions for instances with 5 and 10 customers. For the instances with 15 customers, CPLEX optimally solved only 2 out of 12 cases. For the remaining instances, it produced bounded solutions with an average optimality gap of 7.1\%. Additionally, as the problem size increased, the computational time required by CPLEX grew substantially. Specifically, while CPLEX took an average of 2.72 seconds to optimally solve instances with 5 customers, this increased to 285.26 seconds for instances with 10 customers and further rose to (on average) 3399.92 seconds for instances with 15 customers.

In contrast, the heuristic approaches, including the probabilistic constructive heuristics and CMSA, delivered good-quality solutions in significantly shorter computation times. However, their solutions were generally of slightly lower quality than those obtained by CPLEX. This emphasizes the importance of fine-tuning these algorithms, particularly on smaller problem instances, to achieve better performance.

\begin{table}[h]
\centering
\caption{Computational results for the 2E-EVRP, small-sized instances with 5 customers.}
\label{base_tab:results1}
\scalebox{0.75}{
\begin{tabular}{c|ccc|ccc|ccc} \noalign{\hrule height 1pt}
\multicolumn{1}{c|}{\textbf{Instances}} & \multicolumn{3}{c|}{\textbf{CPLEX}} & \multicolumn{3}{c|}{\boldmath{\textbf{CONSTRUCTIVE HEURISTICS}}} & \multicolumn{3}{c}{\boldmath{\textbf{CMSA}}} \\
\textbf{Name} & \textbf{Dist} &  \boldmath{$\overline{t(s)}$}&  \textbf{Gap(\%)}  & \textbf{Dist} & \textbf{Avg} & \boldmath{$\overline{t(s)}$} & \textbf{Dist} & \textbf{Avg} & \boldmath{$\overline{t(s)}$} \\
\hline
\texttt{C101\_C5x} & \bf325.0 & 0.35 & 0.0 & \bf325.0 & 325.0 & 0.0 & \bf325.0 & 325.0 & 0.0 \\
\texttt{C103\_C5x} & \bf298.0 & 0.47 & 0.0 & \bf298.0 & 298.0 & 0.0 & \bf298.0 & 298.0 & 0.0 \\
\texttt{C206\_C5x} & \bf351.0 & 1.46 & 0.0 & \bf351.0 & 351.0 & 0.0 & \bf351.0 & 351.0 & 0.0 \\
\texttt{C208\_C5x} & \bf366.0 & 1.34 & 0.0 & 382.0 & 382.0 & 0.11 & 382.0 & 382.0 & 0.03 \\
\texttt{R104\_C5x} & \bf316.0 & 1.32 & 0.0 & \bf316.0 & 316.0 & 0.0 & \bf316.0 & 316.0 & 0.0 \\
\texttt{R105\_C5x} & \bf352.0 & 0.74 & 0.0 & \bf372.0 & 372.0 & 0.0 & \bf352.0 & 352.0 & 0.02 \\
\texttt{R202\_C5x} & \bf348.0 & 2.61 & 0.0 & \bf348.0 & 348.0 & 0.01 & \bf348.0 & 348.0 & 0.0 \\
\texttt{R203\_C5x} & \bf372.0 & 5.38 & 0.0 & \bf372.0 & 372.0 & 0.0 & \bf372.0 & 372.0 & 0.0 \\
\texttt{RC105\_C5x} & \bf356.0 & 4.25 & 0.0 & 407.0 & 407.0 & 0.0 & 407.0 & 407.0 & 0.0 \\
\texttt{RC108\_C5x} & \bf380.0 & 11.72 & 0.0 & 430.0 & 430.0 & 0.01 & 430.0 & 430.0 & 0.01 \\
\texttt{RC204\_C5x} & \bf332.0 & 2.05 & 0.0 & 387.0 & 387.0 & 0.0 & \bf332.0 & 332.0 & 0.02 \\
\texttt{RC208\_C5x} & \bf328.0 & 1.0 & 0.0 & \bf328.0 & 328.0 & 0.0 & \bf328.0 & 328.0 & 0.0 \\
\hline \hline
\texttt{average} & \bf343.67 & 2.72 & 0.0 & 359.67 & 359.67 & 0.01 & 353.42 & 353.42 & 0.01 \\

\hline
\end{tabular}}
\end{table}

\begin{table}[h]
\centering
\caption{Computational results for the 2E-EVRP, small-sized instances with 10 customers.}
\label{base_tab:results2}
\scalebox{0.75}{
\begin{tabular}{c|ccc|ccc|ccc} \noalign{\hrule height 1pt}
\multicolumn{1}{c|}{\textbf{Instances}} & \multicolumn{3}{c|}{\textbf{CPLEX}} & \multicolumn{3}{c|}{\boldmath{\textbf{CONSTRUCTIVE HEURISTICS}}} & \multicolumn{3}{c}{\boldmath{\textbf{CMSA}}} \\
\textbf{Name} & \textbf{Dist} &  \boldmath{$\overline{t(s)}$}&  \textbf{Gap(\%)}  & \textbf{Dist} & \textbf{Avg} & \boldmath{$\overline{t(s)}$} & \textbf{Dist} & \textbf{Avg} & \boldmath{$\overline{t(s)}$} \\
\hline
\texttt{C101\_C10x} & \bf471.0 & 1175.67 & 0.0 & 486.0 & 486.0 & 1.17 & 485.0 & 485.0 & 0.16 \\
\texttt{C104\_C10x} & \bf413.0 & 10.46 & 0.0 & 485.0 & 485.0 & 3.32 & 485.0 & 485.0 & 0.06 \\
\texttt{C202\_C10x} & \bf369.0 & 92.48 & 0.0 & 406.0 & 406.0 & 3.66 & 406.0 & 406.0 & 0.05 \\
\texttt{C205\_C10x} & \bf402.0 & 139.21 & 0.0 & 416.0 & 418.7 & 1.26 & 404.0 & 404.0 & 0.29 \\
\texttt{R102\_C10x} & \bf359.0 & 19.14 & 0.0 & 441.0 & 447.8 & 69.97 & 440.0 & 440.0 & 0.05 \\
\texttt{R103\_C10x} & \bf330.0 & 20.33 & 0.0 & 370.0 & 370.0 & 12.86 & 370.0 & 370.0 & 0.07 \\
\texttt{R201\_C10x} & \bf349.0 & 36.29 & 0.0 & \bf349.0 & 349.0 & 0.01 & \bf349.0 & 349.0 & 0.09 \\
\texttt{R203\_C10x} & \bf436.0 & 1494.21 & 0.0 & \bf436.0 & 436.3 & 29.84 & \bf436.0 & 436.0 & 0.18 \\
\texttt{RC102\_C10x} & \bf455.0 & 20.21 & 0.0 & 534.0 & 534.0 & 1.95 & 534.0 & 534.0 & 0.08 \\
\texttt{RC108\_C10x} & \bf472.0 & 180.62 & 0.0 & 508.0 & 508.0 & 6.07 & 508.0 & 508.0 & 0.15 \\
\texttt{RC201\_C10x} & \bf395.0 & 109.05 & 0.0 & \bf395.0 & 395.0 & 0.74 & \bf395.0 & 395.0 & 0.14 \\
\texttt{RC205\_C10x} & \bf487.0 & 125.46 & 0.0 & \bf487.0 & 487.0 & 30.87 & \bf487.0 & 487.0 & 0.14 \\
\hline \hline
\texttt{average} & 411.5 & 285.26 & 0.0 & 442.75 & 443.57 & 13.48 & 441.58 & 441.58 & 0.12 \\

\hline
\end{tabular}}
\end{table}

\begin{table}[h]
\centering
\caption{Computational results for the 2E-EVRP, small-sized instances with 15 customers.}
\label{base_tab:results3}
\scalebox{0.75}{
\begin{tabular}{c|ccc|ccc|ccc} \noalign{\hrule height 1pt}
\multicolumn{1}{c|}{\textbf{Instances}} & \multicolumn{3}{c|}{\textbf{CPLEX}} & \multicolumn{3}{c|}{\boldmath{\textbf{CONSTRUCTIVE HEURISTICS}}} & \multicolumn{3}{c}{\boldmath{\textbf{CMSA}}} \\
\textbf{Name} & \textbf{Dist} &  \boldmath{$\overline{t(s)}$}&  \textbf{Gap(\%)}  & \textbf{Dist} & \textbf{Avg} & \boldmath{$\overline{t(s)}$} & \textbf{Dist} & \textbf{Avg} & \boldmath{$\overline{t(s)}$} \\
\hline
\texttt{C103\_C15x} & \bf400.0 & 2869.01 & 0.0 & 522.0 & 522.0 & 3.68 & 522.0 & 522.0 & 0.05 \\
\texttt{C106\_C15x} & \bf407.0 & 3591.47 & 9.32 & 446.0 & 508.1 & 20.45 & 424.0 & 424.0 & 2.24 \\
\texttt{C202\_C15x} & \bf460.0 & 3591.22 & 3.94 & 520.0 & 528.8 & 68.24 & 508.0 & 508.0 & 8.76 \\
\texttt{C208\_C15x} & \bf449.0 & 3589.81 & 8.18 & 526.0 & 526.0 & 76.57 & 502.0 & 502.0 & 1.32 \\
\texttt{R102\_C15x} & \bf450.0 & 3591.52 & 15.42 & 535.0 & 538.0 & 23.73 & 497.0 & 497.0 & 11.79 \\
\texttt{R105\_C15x} & \bf449.0 & 3591.5 & 17.12 & 547.0 & 551.9 & 38.11 & 547.0 & 547.0 & 0.17 \\
\texttt{R202\_C15x} & \bf429.0 & 3591.51 & 2.9 & 518.0 & 531.3 & 26.38 & 485.0 & 485.0 & 26.61 \\
\texttt{R209\_C15x} & \bf409.0 & 2017.69 & 0.0 & \bf409.0 & 409.5 & 17.58 & \bf409.0 & 409.0 & 0.6 \\
\texttt{RC103\_C15x} & \bf447.0 & 3591.52 & 5.14 & 589.0 & 592.8 & 68.47 & 505.0 & 505.0 & 9.76 \\
\texttt{RC108\_C15x} & \bf470.0 & 3591.0 & 7.75 & 588.0 & 588.0 & 6.25 & 588.0 & 588.0 & 0.07 \\
\texttt{RC202\_C15x} & \bf454.0 & 3591.25 & 4.57 & 455.0 & 455.0 & 103.01 & \bf454.0 & 454.0 & 0.58 \\
\texttt{RC204\_C15x} & \bf435.0 & 3591.56 & 10.84 & 549.0 & 549.0 & 66.11 & 483.0 & 483.0 & 10.28 \\
\hline \hline
\texttt{average} & \bf438.25 & 3399.92 & 7.1 & 517.0 & 525.03 & 43.22 & 493.67 & 493.67 & 6.02 \\

\hline
\end{tabular}}
\end{table}

For the larger-sized instances, CPLEX was unable to produce valid solutions within the 2-hour computational time limit. Consequently, we compared only the constructive heuristics and CMSA. The results presented in Table~\ref{base_tab:results4} highlight the clear advantage of CMSA over the constructive heuristics in terms of both solution quality and computational efficiency. This demonstrates the robustness and scalability of CMSA, especially when dealing with larger problem sizes.

\begin{table}[h]
\centering
\caption{Computational results for the 2E-EVRP, large-sized instances with 100 customers.}
\label{base_tab:results4}
\scalebox{0.75}{
\begin{tabular}{c|ccc|ccc} \noalign{\hrule height 1pt}
\multicolumn{1}{c|}{\textbf{Instances}} & \multicolumn{3}{c|}{\boldmath{\textbf{CONSTRUCTIVE HEURISTICS}}} & \multicolumn{3}{c}{\boldmath{\textbf{CMSA}}} \\
\textbf{Name}  & \textbf{Dist} & \textbf{Avg} & \boldmath{$\overline{t(s)}$} & \textbf{Dist} & \textbf{Avg} & \boldmath{$\overline{t(s)}$} \\
\hline
\texttt{C101\_21x} & 1594.0 & 1614.7 & 422.94 & \bf1391.0 & \bf1392.4 & 410.54 \\
\texttt{C102\_21x} & 1562.0 & 1602.6 & 635.45 & \bf1391.0 & \bf1391.7 & 346.25 \\
\texttt{C103\_21x} & 1583.0 & 1602.8 & 879.25 & \bf1384.0 & \bf1390.6 & 347.93 \\
\texttt{C104\_21x} & 1594.0 & 1600.9 & 698.53 & \bf1383.0 & \bf1388.0 & 421.89 \\
\texttt{C105\_21x} & 1599.0 & 1601.5 & 592.56 & \bf1391.0 & \bf1391.6 & 239.81 \\
\texttt{C106\_21x} & 1604.0 & 1605.0 & 642.44 & \bf1381.0 & \bf1390.2 & 404.58 \\
\texttt{C107\_21x} & 1578.0 & 1605.9 & 304.57 & \bf1391.0 & \bf1391.5 & 405.1 \\
\texttt{C108\_21x} & 1580.0 & 1604.5 & 282.81 & \bf1391.0 & \bf1391.3 & 473.74 \\
\texttt{C109\_21x} & 1615.0 & 1620.4 & 784.15 & \bf1391.0 & \bf1391.0 & 274.23 \\
\texttt{C201\_21x} & 1323.0 & 1324.7 & 351.64 & \bf1123.0 & \bf1132.0 & 602.7 \\
\texttt{C202\_21x} & 1323.0 & 1342.5 & 580.24 & \bf1122.0 & \bf1128.5 & 421.78 \\
\texttt{C203\_21x} & 1342.0 & 1350.1 & 352.7 & \bf1129.0 & \bf1139.1 & 412.53 \\
\texttt{C204\_21x} & 1287.0 & 1322.4 & 276.02 & \bf1122.0 & \bf1134.6 & 514.66 \\
\texttt{C205\_21x} & 1264.0 & 1316.5 & 575.4 & \bf1126.0 & \bf1138.2 & 339.18 \\
\texttt{C206\_21x} & 1283.0 & 1335.6 & 658.83 & \bf1122.0 & \bf1132.4 & 437.23 \\
\texttt{C207\_21x} & 1331.0 & 1349.9 & 455.02 & \bf1122.0 & \bf1132.8 & 491.97 \\
\texttt{C208\_21x} & 1349.0 & 1349.3 & 652.56 & \bf1126.0 & \bf1131.5 & 458.39 \\
\texttt{R101\_21x} & 1592.0 & 1607.7 & 716.32 & \bf1341.0 & \bf1352.0 & 435.84 \\
\texttt{R102\_21x} & 1555.0 & 1562.5 & 748.0 & 1\bf335.0 & \bf1347.2 & 612.21 \\
\texttt{R103\_21x} & 1555.0 & 1573.1 & 524.54 & \bf1343.0 & \bf1347.7 & 543.8 \\
\texttt{R104\_21x} & 1561.0 & 1569.7 & 343.94 & \bf1331.0 & \bf1339.0 & 435.32 \\
\texttt{R105\_21x} & 1573.0 & 1599.3 & 418.1 & \bf1338.0 & \bf1346.5 & 633.02 \\
\texttt{R106\_21x} & 1478.0 & 1516.0 & 129.01 & \bf1338.0 & \bf1342.5 & 433.78 \\
\texttt{R107\_21x} & 1466.0 & 1576.7 & 611.63 & \bf1330.0 & \bf1341.5 & 701.07 \\
\texttt{R108\_21x} & 1529.0 & 1547.6 & 614.42 & \bf1343.0 & \bf1354.7 & 499.56 \\
\texttt{R109\_21x} & 1529.0 & 1562.0 & 648.33 & \bf1330.0 & \bf1335.1 & 617.68 \\
\texttt{R110\_21x} & 1562.0 & 1585.5 & 302.54 & \bf1330.0 & \bf1335.9 & 343.31 \\
\texttt{R111\_21x} & 1483.0 & 1583.4 & 467.91 & \bf1332.0 & \bf1344.0 & 593.16 \\
\texttt{R112\_21x} & 1573.0 & 1634.4 & 454.68 & \bf1332.0 & \bf1344.2 & 551.44 \\
\texttt{R201\_21x} & 1108.0 & 1149.6 & 187.74 & \bf941.0 & \bf945.5 & 504.99 \\
\texttt{R202\_21x} & 1107.0 & 1107.0 & 482.76 & \bf939.0 & \bf947.6 & 325.04 \\
\texttt{R203\_21x} & 1108.0 & 1137.0 & 585.65 & \bf939.0 & \bf947.6 & 594.07 \\
\texttt{R204\_21x} & 1080.0 & 1138.3 & 631.39 & \bf937.0 & \bf943.8 & 667.1 \\
\texttt{R205\_21x} & 1111.0 & 1137.7 & 617.89 & \bf940.0 & \bf946.3 & 478.31 \\
\texttt{R206\_21x} & 1084.0 & 1102.4 & 334.15 & \bf945.0 & \bf949.5 & 342.9 \\
\texttt{R207\_21x} & 1086.0 & 1118.1 & 458.22 & \bf940.0 & \bf945.9 & 359.82 \\
\texttt{R208\_21x} & 1054.0 & 1096.7 & 592.02 & \bf941.0 & \bf949.9 & 411.23 \\
\texttt{R209\_21x} & 1076.0 & 1150.4 & 649.79 & \bf940.0 & \bf944.8 & 451.09 \\
\texttt{R210\_21x} & 1092.0 & 1132.0 & 725.88 & \bf942.0 & \bf949.0 & 372.89 \\
\texttt{R211\_21x} & 1075.0 & 1081.8 & 865.61 & \bf937.0 & \bf946.2 & 484.5 \\
\texttt{RC101\_21x} & 1791.0 & 1886.2 & 637.65 & \bf1603.0 & \bf1604.4 & 302.16 \\
\texttt{RC102\_21x} & 1841.0 & 1915.7 & 393.94 & \bf1603.0 & \bf1605.1 & 438.36 \\
\texttt{RC103\_21x} & 1794.0 & 1928.5 & 223.17 & \bf1603.0 & \bf1603.9 & 316.51 \\
\texttt{RC104\_21x} & 1827.0 & 1913.4 & 534.34 & \bf1603.0 & \bf1603.9 & 383.2 \\
\texttt{RC105\_21x} & 1899.0 & 1908.2 & 555.59 & \bf1603.0 & \bf1604.8 & 371.8 \\
\texttt{RC106\_21x} & 1901.0 & 1903.8 & 503.66 & \bf1603.0 & \bf1603.5 & 340.07 \\
\texttt{RC107\_21x} & 1892.0 & 1910.8 & 603.32 & \bf1603.0 & \bf1603.5 & 335.08 \\
\texttt{RC108\_21x} & 1884.0 & 1895.4 & 556.27 & \bf1603.0 & \bf1605.1 & 365.53 \\
\texttt{RC201\_21x} & 1125.0 & 1156.0 & 842.06 & \bf982.0 & \bf991.1 & 585.8 \\
\texttt{RC202\_21x} & 1093.0 & 1132.3 & 615.67 & \bf988.0 & \bf993.5 & 226.25 \\
\texttt{RC203\_21x} & 1110.0 & 1141.4 & 757.88 & \bf979.0 & \bf992.7 & 378.34 \\
\texttt{RC204\_21x} & 1108.0 & 1163.7 & 662.59 & \bf985.0 & \bf993.7 & 502.36 \\
\texttt{RC205\_21x} & 1115.0 & 1156.0 & 748.07 & \bf986.0 & \bf994.2 & 486.5 \\
\texttt{RC206\_21x} & 1114.0 & 1163.8 & 572.47 & \bf987.0 & \bf991.8 & 506.18 \\
\texttt{RC207\_21x} & 1169.0 & 1185.8 & 663.33 & \bf987.0 & \bf991.6 & 308.85 \\
\texttt{RC208\_21x} & 1153.0 & 1170.0 & 545.68 & \bf987.0 & \bf991.7 & 425.65 \\
\hline \hline
\texttt{average} & 1411.86 & 1445.52 & 547.7 & \bf1224.2 & \bf1230.51 & 440.49 \\

\hline
\end{tabular}}
\end{table}

\subsection{Two Echelon Electric Vehicle Routing Problem with Time Windows and Satellite Synchronization}

In logistics, time windows are a crucial aspect of real-world delivery problems. Time windows ensure that deliveries or pickups are completed within specific time intervals, thus improving customer satisfaction and reducing logistical inefficiencies. In variants of the Two Echelon Electric Vehicle Routing Problem with Time Windows (2E-EVRP-TW), only the second-echelon nodes (i.e., customers) are constrained by time windows. These time windows are defined by the earliest and latest times that a customer can be serviced. The goal is to ensure that electric vehicles arrive at the customer locations within their designated time windows while considering other constraints such as battery consumption and satellite synchronization.

\subsubsection{Notation}
\begin{itemize}
    \item $e_i$: earliest arrival time for the second echelon nodes $i \in N_{SRCS}$.
    \item $l_i$: latest arrival time for the second echelon nodes $i \in N_{SRCS}$.
    \item $t_{ij}$: travel time between node $i$ and node $j$. 
    \item $s_i$: service time at node $i$.
    \item $\tau_i$: arrival time at node $i$.
    \item $M$: a sufficiently large constant.
\end{itemize}

\subsubsection{Additional Constraints for Time Windows and Satellite Synchronization}

To incorporate time windows into the model, the following constraints are added to the base 2E-EVRP model presented in Equations~(\ref{m1:1})-(\ref{m1:22}). These constraints ensure that vehicles respect time window constraints in the second echelon and synchronize with satellite operations.

\begin{align}
    & \tau_{i}^{1} = 0, & \forall i \in N_D \label{tw1}
\end{align}
This constraint initializes the time for large trucks in the first echelon. Central depots are assumed to be the starting point, and time starts at zero.

\begin{align}
    & \tau_{i}^{1} + (t_{ij}^{1} + s_i) x_{ij} - M(1 - x_{ij}) \leq \tau_{j}^{1}, & \forall i \in N_{DS}, \forall j \in N_{SD}, i \neq j \label{tw2}
\end{align}
This constraint calculates the arrival time at each satellite, ensuring that large trucks respect the service and travel times between nodes in the first echelon.

\begin{align}
    & \tau_{j}^{2} \geq \tau_{j}^{1}, & \forall j \in N_S \label{tw3}
\end{align}
This ensures that electric vehicles (second echelon) cannot start servicing a satellite before a large truck (first echelon) has finished its operations at the same satellite, maintaining synchronization between the two fleets.

\begin{align}
    & \tau_{i}^{2} + (t_{ij}^{2} + s_i) y_{ij} - M(1 - y_{ij}) \leq \tau_{j}^{2}, & \forall i \in N_{SRC}, \forall j \in N_{RCS}, i \neq j \label{tw4}
\end{align}
This constraint ensures that electric vehicles respect the travel and service times between nodes in the second echelon, including satellites, charging stations, and customers.

\begin{align}
    & \tau_{i}^{2} + t_{ij}^{2} y_{ij} + g(B - BSCa_i) - (M + gQ)(1 - y_{ij}) \leq \tau_{j}^{2}, & \forall i \in N_R, \forall j \in N_{RCS}, i \neq j \label{tw5}
\end{align}
This constraint accounts for battery charging time at charging stations. If the electric vehicle visits a charging station, the arrival time at the next node must reflect the charging duration.

\begin{align}
    & e_j \leq \tau_j^{2} \leq l_j, & \forall j \in N_C \label{tw6}
\end{align}
This constraint ensures that electric vehicles visit customers within their specified time windows. The variable $\tau_j^{2}$ must be within the range $[e_j, l_j]$, where $e_j$ is the earliest time and $l_j$ is the latest time for service at customer $j$.

\subsubsection{Computational Results}

In the following we provide a detailed comparison of the following methods. First, we applied both CPLEX (version 12.10) and our Clarke-Wright Savings Heuristic to all problem instances. Hereby, CPLEX was given a time limit of 2 h of CPU time for each problem instance. Next, we also applied two versions of VNS from~\cite{akbay2022variable}, \VNSred\ and \VNSfull, extended for respecting satellite synchronization. Both versions of VNS were applied with a computation time limit of 150 CPU seconds in the case of small problem instances, and 900 CPU seconds for large problem instances. Moreover, both versions of VNS were applied 10 times to each problem instance. 

Tables~\ref{tw:results1}--\ref{tw:results3} show the numerical results for small problem instances with 5, 10, and 15~customers, respectively. The structure of these tables is as follows. Instance names are given in the first column. After the first table column, there are four blocks of columns, presenting the results of our four approaches. The first column of each block (with headings 'Dist') is the same for all four approaches. Hereby column 'Dist' provides the objective function values of the solutions generated by the four approaches. In the case of \VNSfull\ and \VNSred, 'Dist' shows the objective function value of the best solution found in 10 runs, while an additional column with the heading 'Avg' provides the average objective function value of the best solutions of each of the 10 runs. Next, columns with heading '$\overline{t(s)}$' show the computation time of CPLEX, our Clarke-Wright Savings Heuristic, and the average computation times of \VNSfull\ and \VNSred\ to find the best solutions in each run. Finally, column 'Gap(\%)' provides the gap (in percent) between the best solution and the best lower bound found by CPLEX. Note that, in the case where the gap value is zero, CPLEX has found an optimal solution.

\renewcommand{\arraystretch}{1.1}																							
\begin{table}[!t]																							
\centering																							
\caption{Computational results for the 2E-EVRP-TW-SS, small-sized instances with 5 customers.}																							
\label{tw:results1}																							
\scalebox{0.75}{																							
\begin{tabular}{c|ccc|cc|ccc|ccc} \noalign{\hrule height 1pt} 																							
\multicolumn{1}{c|}{\textbf{Instances}}      & \multicolumn{3}{c|}{\textbf{CPLEX}}   & \multicolumn{2}{c|}{\textbf{Clarke-Wright Savings Heuristic}} & \multicolumn{3}{c|}{\boldmath{\textbf{\VNSred}}}    & \multicolumn{3}{c}{\boldmath{\textbf{\VNSfull}}} \\ 																							
\textbf{Name}	&	\textbf{Dist}	&	\textbf{Gap(\%)}	&\boldmath{$\overline{t(s)}$}  &	\textbf{Dist}	&	\boldmath{$\overline{t(s)}$	}&	\textbf{Dist}	&	\textbf{Avg}	&	\boldmath{$\overline{t(s)}$}	&	\textbf{Dist}	&	\textbf{Avg}	&	\boldmath{$\overline{t(s)}$}	\\	\hline	
\texttt{C101\_C5}	&	\bf385.49	&	0	&	1.67	&	442.19	&	0.00021	&	\bf385.49	&	\bf385.49	&	0.989	&	\bf385.49	&	\bf385.49	&	12.509	\\
\texttt{C103\_C5}	&	\bf341.33	&	0	&	0.09	&	360.94	&	0.00011	&	\bf341.33	&	\bf341.33	&	0.006	&	\bf341.33	&	\bf341.33	&	0.502	\\
\texttt{C206\_C5}	&	\bf417.31	&	0	&	5.97	&	480.9	&	0.00017	&	\bf417.31	&	\bf417.31	&	0.001	&	\bf417.31	&	\bf417.31	&	0.001	\\
\texttt{C208\_C5}	&	\bf381.91	&	0	&	0.31	&	383.07	&	0.00011	&	\bf381.91	&	\bf381.91	&	0.001	&	\bf381.91	&	\bf381.91	&	0.001	\\
\texttt{R104\_C5}	&	\bf317.02	&	0	&	1.61	&	317.78	&	0.00012	&	\bf317.02	&	\bf317.02	&	0.001	&	\bf317.02	&	\bf317.02	&	0.001	\\
\texttt{R105\_C5}	&	\bf453.74	&	0	&	9.57	&	677.61	&	0.00014	&	\bf453.74	&	495.16	&	0	&	\bf453.74	&	\bf453.74	&	29.693	\\
\texttt{R202\_C5}	&	\bf347.82	&	0	&	0.21	&	348.29	&	0.0001	&	\bf347.82	&	\bf347.82	&	0.001	&	\bf347.82	&	\bf347.82	&	0.001	\\
\texttt{R203\_C5}	&	\bf371.31	&	0	&	0.21	&	387.92	&	0.00016	&	386.48	&	386.48	&	0.001	&	\bf371.31	&	\bf371.31	&	7.203	\\
\texttt{RC105\_C5}	&	\bf432.64	&	0	&	28.84	&	496.72	&	0.00015	&	\bf432.64	&	435.77	&	0.404	&	\bf432.64	&	437.34	&	21.488	\\
\texttt{RC108\_C5}	&	\bf460.89	&	0	&	24.24	&	702.23	&	0.00016	&	\bf460.89	&	\bf460.89	&	0.008	&	\bf460.89	&	\bf460.89	&	3.281	\\
\texttt{RC204\_C5}	&	\bf332.86	&	0	&	0.64	&	649.44	&	0.00015	&	\bf332.86	&	\bf332.86	&	0.018	&	\bf332.86	&	\bf332.86	&	0.015	\\
\texttt{RC208\_C5}	&	\bf327.30	&	0	&	0.37	&	331.77	&	0.0001	&	331.77	&	331.77	&	0	&	\bf327.30	&	\bf327.30	&	15.193	\\
\hline																							
\hline																							
\textbf{average}	&	\bf380.80	&	-	&	6.15	&	464.905	&	0.00014	&	382.44	&	386.15	&	0.119	&	\bf380.80	&	381.19	&	7.491	\\
\hline																							
																							
\end{tabular}}																							
\end{table}

\renewcommand{\arraystretch}{1.1}																							
\begin{table}[!t]																							
\centering																							
\caption{Computational results for the 2E-EVRP-TW-SS, small-sized instances with 10 customers.}																							
\label{tw:results2}																							
\scalebox{0.75}{																							
\begin{tabular}{c|ccc|cc|ccc|ccc} \noalign{\hrule height 1pt} 																							
\multicolumn{1}{c|}{\textbf{Instances}}      & \multicolumn{3}{c|}{\textbf{CPLEX}}   & \multicolumn{2}{c|}{\textbf{Clarke-Wright Savings Heuristic}} & \multicolumn{3}{c|}{\boldmath{\textbf{\VNSred}}}    & \multicolumn{3}{c}{\boldmath{\textbf{\VNSfull}}} \\ 																							
\textbf{Name}	&	\textbf{Dist}	&	\textbf{Gap(\%)}	&\boldmath{$\overline{t(s)}$}  &	\textbf{Dist}	&	\boldmath{$\overline{t(s)}$	}&	\textbf{Dist}	&	\textbf{Avg}	&	\boldmath{$\overline{t(s)}$}	&	\textbf{Dist}	&	\textbf{Avg}	&	\boldmath{$\overline{t(s)}$}	\\	\hline	
\texttt{C103\_C15}	&	-	&	-	&	-	&	690.99	&	0.00036	&	\bf575.18	&	582.02	&	4.925	&	\bf575.18	&	\bf575.18	&	0.623	\\
\texttt{C106\_C15}	&	\bf500.32	&	13.37	&	7182.91	&	681.31	&	0.00022	&	516.6	&	524.1	&	2.1	&	516.6	&	516.6	&	1.027	\\
\texttt{C202\_C15}	&	714.81	&	32.23	&	7183.04	&	729.87	&	0.00034	&	617.24	&	618.66	&	29.966	&	\bf550.32	&	\bf550.32	&	12.454	\\
\texttt{C208\_C15}	&	\bf550.02	&	15.56	&	7182.95	&	737.61	&	0.00023	&	619.73	&	619.73	&	6.976	&	\bf550.02	&	\bf550.02	&	22	\\
\texttt{R102\_C15}	&	-	&	-	&	-	&	950.25	&	0.00026	&	\bf716.56	&	\bf716.56	&	9.523	&	\bf716.56	&	\bf716.56	&	12.056	\\
\texttt{R105\_C15}	&	-	&	-	&	-	&	777.77	&	0.00038	&	\bf607.96	&	\bf607.96	&	30.85	&	\bf607.96	&	\bf607.96	&	25.605	\\
\texttt{R202\_C15}	&	719.61	&	35.36	&	7198.17	&	990.37	&	0.00043	&	\bf593.69	&	597.79	&	8.033	&	\bf593.69	&	\bf593.69	&	60.988	\\
\texttt{R209\_C15}	&	\bf475.10	&	10.09	&	7182.43	&	711.09	&	0.00024	&	\bf475.10	&	519.46	&	0.712	&	\bf475.10	&	482.3	&	77.386	\\
\texttt{RC103\_C15}	&	-	&	-	&	-	&	745.82	&	0.00035	&	\bf616.32	&	622.1	&	1.565	&	\bf616.32	&	\bf616.32	&	1.803	\\
\texttt{RC108\_C15}	&	-	&	-	&	-	&	716.22	&	0.00026	&	\bf603.87	&	\bf603.87	&	0.214	&	\bf603.87	&	615.11	&	0.033	\\
\texttt{RC202\_C15}	&	\bf552.70	&	16.06	&	7182.65	&	697.24	&	0.00033	&	601.86	&	601.86	&	2.395	&	\bf552.70	&	587.11	&	11.6	\\
\texttt{RC204\_C15}	&	\bf485.34	&	13.93	&	7183.03	&	604.05	&	0.00035	&	551.56	&	551.56	&	0.67	&	\bf485.34	&	\bf485.34	&	15.566	\\
																							
\hline \hline																							
\textbf{average}	&	-	&	-	&	-	&	752.72	&	0.00031	&	591.31	&	597.14	&	8.161	&	\bf570.30	&	574.71	&	20.095	\\
\hline																							
\end{tabular}}																							
\end{table}

\renewcommand{\arraystretch}{1.1}																							
\begin{table}[!t]																							
\centering																							
\caption{Computational results for the 2E-EVRP-TW-SS, small-sized instances with 15 customers.}																							
\label{tw:results3}																							
\scalebox{0.75}{																							
\begin{tabular}{c|ccc|cc|ccc|ccc} \noalign{\hrule height 1pt} 																							
\multicolumn{1}{c|}{\textbf{Instances}}      & \multicolumn{3}{c|}{\textbf{CPLEX}}   & \multicolumn{2}{c|}{\textbf{Clarke-Wright Savings Heuristic}} & \multicolumn{3}{c|}{\boldmath{\textbf{\VNSred}}}    & \multicolumn{3}{c}{\boldmath{\textbf{\VNSfull}}} \\ 																							
\textbf{Name}	&	\textbf{Dist}	&	\textbf{Gap(\%)}	&\boldmath{$\overline{t(s)}$}  &	\textbf{Dist}	&	\boldmath{$\overline{t(s)}$	}&	\textbf{Dist}	&	\textbf{Avg}	&	\boldmath{$\overline{t(s)}$}	&	\textbf{Dist}	&	\textbf{Avg}	&	\boldmath{$\overline{t(s)}$}	\\	\hline	
\texttt{C103\_C15}	&	-	&	-	&	-	&	690.99	&	0.00036	&	\bf575.18	&	582.02	&	4.925	&	\bf575.18	&	\bf575.18	&	0.623	\\
\texttt{C106\_C15}	&	\bf500.32	&	13.37	&	7182.91	&	681.31	&	0.00022	&	516.6	&	524.1	&	2.1	&	516.6	&	516.6	&	1.027	\\
\texttt{C202\_C15}	&	714.81	&	32.23	&	7183.04	&	729.87	&	0.00034	&	617.24	&	618.66	&	29.966	&	\bf550.32	&	\bf550.32	&	12.454	\\
\texttt{C208\_C15}	&	\bf550.02	&	15.56	&	7182.95	&	737.61	&	0.00023	&	619.73	&	619.73	&	6.976	&	\bf550.02	&	\bf550.02	&	22	\\
\texttt{R102\_C15}	&	-	&	-	&	-	&	950.25	&	0.00026	&	\bf716.56	&	\bf716.56	&	9.523	&	\bf716.56	&	\bf716.56	&	12.056	\\
\texttt{R105\_C15}	&	-	&	-	&	-	&	777.77	&	0.00038	&	\bf607.96	&	\bf607.96	&	30.85	&	\bf607.96	&	\bf607.96	&	25.605	\\
\texttt{R202\_C15}	&	719.61	&	35.36	&	7198.17	&	990.37	&	0.00043	&	\bf593.69	&	597.79	&	8.033	&	\bf593.69	&	\bf593.69	&	60.988	\\
\texttt{R209\_C15}	&	\bf475.10	&	10.09	&	7182.43	&	711.09	&	0.00024	&	\bf475.10	&	519.46	&	0.712	&	\bf475.10	&	482.3	&	77.386	\\
\texttt{RC103\_C15}	&	-	&	-	&	-	&	745.82	&	0.00035	&	\bf616.32	&	622.1	&	1.565	&	\bf616.32	&	\bf616.32	&	1.803	\\
\texttt{RC108\_C15}	&	-	&	-	&	-	&	716.22	&	0.00026	&	\bf603.87	&	\bf603.87	&	0.214	&	\bf603.87	&	615.11	&	0.033	\\
\texttt{RC202\_C15}	&	\bf552.70	&	16.06	&	7182.65	&	697.24	&	0.00033	&	601.86	&	601.86	&	2.395	&	\bf552.70	&	587.11	&	11.6	\\
\texttt{RC204\_C15}	&	\bf485.34	&	13.93	&	7183.03	&	604.05	&	0.00035	&	551.56	&	551.56	&	0.67	&	\bf485.34	&	\bf485.34	&	15.566	\\
																							
\hline \hline																							
\textbf{average}	&	-	&	-	&	-	&	752.72	&	0.00031	&	591.31	&	597.14	&	8.161	&	\bf570.30	&	574.71	&	20.095	\\
\hline																							
\end{tabular}}																							
\end{table}

The following observations can be made: First, apart from instances \texttt{R103\_C10}, \texttt{RC102\_C10}, and \texttt{RC108\_C10}, CPLEX was able to solve the mathematical model---within 2 h of CPU time---for all instances with five and ten customers to optimality. For two of the remaining three cases, CPLEX was able to provide feasible solutions of the same quality as \VNSfull\ and \VNSred, without being able to prove optimality. However, for the instances with 15 customers, the performance of CPLEX heavily starts to degrade. The reason for the rapidly decreasing performance of CPLEX is that the size and complexity of the MILP model sharply increase based on the instance size. For instance, the average number of variables and constraints of the MILP model for the instances containing five customers is 986 and 2235, respectively. These values increase to 4008 and 9363 for the instances with 10 customers and to 13125 and 31482 for the instances with 15 customers. In this latter case, CPLEX could only provide valid solutions (without being able to prove optimality) in seven out of 12 instances.  Nevertheless, for one instance (\texttt{C106\_15}), CPLEX produced a better solution than both VNS variants. 

Both VNS variants performed comparably on small problem instances with 5 and 10 customers. They were able to find solutions with the same objective function values as those of CPLEX. However, the performance of the two VNS variants starts to differ for the instances with 15 customers. While \VNSfull\ provides results at least as good as CPLEX for all instances except for \texttt{C106\_C15}, \VNSred\ only does so in seven out of 12 cases. Considering those instances for which CPLEX was able to obtain a solution, both VNS variants improved the solution quality of CPLEX, on average, by 0.55\% (\VNSred\ ) and 6.86\% (\VNSfull\ ). In fact, \VNSfull\ outperforms \VNSred\ both in terms of best-performance (column 'Dist') and in terms of average performance (column 'Avg'). Note also that the running times of both \VNSfull\ and \VNSred\ are in the order of seconds. While the superiority of both \VNSfull\ and \VNSred\ over CPLEX in terms of CPU time is more significant for the instances with 10 and 15 customers, see Tables~\ref{tw:results2} and \ref{tw:results3}, only \VNSred\ provides better CPU times for the instances with 5 customers. Finally, note that the results of the Clarke-Wright Savings Heuristic are, in the context of these small problem instances, approx.~20\% worse than the best results obtained. This is, however, achieved in very low computation times of a fraction of a second, which shows that our Clarke-Wright Savings Heuristic is a good candidate for producing the initial solutions of VNS. 

Next, we analyze the results of the four approaches when applied to the large problem instances of our benchmark set. These results are shown in Table~\ref{tw:results4}. The structure of this table is slightly different from the one of the previous result tables. First, results of CPLEX are not provided, because CPLEX was not able to generate a single valid solution within 2 h of computation time. Second, the additional column with heading 'Imp(\%)' provides the improvement (in percent) of the VNS variants over the results of the Clarke-Wright Savings Heuristic. 

The following observations can be made. For large clustered instances {and large random instances), \VNSfull\ significantly outperforms \VNSred, both in terms of best-performance and average performance. However, the opposite is generally the case in the context of random-clustered instances. This means that the removal/destroy operators have a rather negative impact on the performance of VNS in these cases. This is most probably due to their elevated computation time requirements. Moreover, the superiority of \VNSfull\ over \VNSred\ is much more pronounced in the context of instances with a long scheduling horizon (\texttt{R2*} \texttt{C2*} and \texttt{RC2*}) compared to the instances with a short scheduling horizon (\texttt{R1*}, \texttt{C1*} and \texttt{RC1*}). Finally, when considering all large instances together, \VNSfull\ significantly outperforms \VNSred. 

When comparing the algorithms in terms of the average computation times required to find the best solutions in a run, it can be seen that \VNSred\ was able to provide solutions in lower CPU times than \VNSfull. We can infer that destroy and repair type operators help to produce better solutions; however, repairing a destroyed solution prolongs the computation time.

\renewcommand{\arraystretch}{1.1}																						
\begin{table}[!t]																						
\centering																						
\caption{Computational results for the 2E-EVRP-TW-SS, large-sized instances with 100 customers.}																						
\label{tw:results4}																						
\scalebox{0.75}{																						
\begin{tabular}{c|cc|cccc|cccc}\noalign{\hrule height 1pt} 																						
\multicolumn{1}{c|}{\textbf{Instances}}      &   \multicolumn{2}{c|}{\textbf{Clarke-Wright Savings Heuristic}} & \multicolumn{4}{c|}{\boldmath{\textbf{\VNSred}}}    & \multicolumn{4}{c}{\boldmath{\textbf{\VNSfull}}} \\ 																						
\textbf{Name}	&	\textbf{Dist}	&	\boldmath{$\overline{t(s)}$}	&	\textbf{Dist}	&	\textbf{Avg}	&	\textbf{Imp(\%)}	&	\boldmath{$\overline{t(s)}$}	&	\textbf{Dist}	&	\textbf{Avg}	&	\textbf{Imp(\%)}	&	\boldmath{$\overline{t(s)}$}	\\	\hline
\texttt{C101\_C21}	&	1941.16	&	0.005	&	1513.91	&	1562.77	&	19.49	&	499.7	&	\bf1494.18	&	\bf1538.74	&	20.73	&	579.98	\\	
\texttt{C102\_C21}	&	1822.02	&	0.005	&	1501.66	&	1506.95	&	17.29	&	537.57	&	\bf1447.86	&	\bf1487.11	&	18.38	&	572.03	\\	
\texttt{C103\_C21}	&	1702.89	&	0.005	&	1447.98	&	1463.34	&	14.07	&	509.37	&	\bf1399.25	&	\bf1425.80	&	16.27	&	656.63	\\	
\texttt{C104\_C21}	&	1580.07	&	0.005	&	1435.04	&	1446.17	&	8.47	&	405.19	&	\bf1400.52	&	\bf1439.76	&	8.88	&	540.97	\\	
\texttt{C105\_C21}	&	1877.85	&	0.005	&	1522.97	&	1541.6	&	17.91	&	359	&	\bf1493.69	&	\bf1521.13	&	19	&	466.13	\\	
\texttt{C106\_C21}	&	1791.74	&	0.004	&	1474.74	&	1491.7	&	16.75	&	361.26	&	\bf1429.75	&	\bf1476.85	&	17.57	&	536.93	\\	
\texttt{C107\_C21}	&	1838.83	&	0.005	&	1499.81	&	1513.36	&	17.7	&	400.85	&	\bf1485.7	&	\bf1513.18	&	17.71	&	582.44	\\	
\texttt{C108\_C21}	&	1687.15	&	0.005	&	1461.25	&	\bf1476.72	&	12.47	&	483	&	\bf1450.96	&	1489.63	&	11.71	&	523.87	\\	
\texttt{C109\_C21}	&	1619.19	&	0.005	&	1447.36	&	1456.9	&	10.02	&	326.59	&	\bf1409.97	&	\bf1455.93	&	10.08	&	673.76	\\	
\texttt{C201\_C21}	&	1794.83	&	0.004	&	1251.62	&	1276.42	&	28.88	&	422.74	&	\bf1208.76	&	\bf1233.87	&	31.25	&	545.02	\\	
\texttt{C202\_C21}	&	1672.52	&	0.005	&	1228.61	&	1260.08	&	24.66	&	532.72	&	\bf1187.87	&	\bf1232.74	&	26.29	&	703.08	\\	
\texttt{C203\_C21}	&	1554.96	&	0.005	&	\bf1197.45	&	1223.16	&	21.34	&	356.27	&	1201.4	&	\bf1216.36	&	21.78	&	767.58	\\	
\texttt{C204\_C21}	&	1411.07	&	0.005	&	1178.14	&	1191.92	&	15.53	&	464.31	&	\bf1161.07	&	\bf1181.40	&	16.28	&	577.96	\\	
\texttt{C205\_C21}	&	1470.73	&	0.005	&	1226.48	&	1249.59	&	15.04	&	460.15	&	\bf1205.23	&	\bf1223.94	&	16.78	&	664.39	\\	
\texttt{C206\_C21}	&	1399.11	&	0.005	&	1202.52	&	1222.94	&	12.59	&	519.78	&	\bf1182.63	&	\bf1198.54	&	14.34	&	556.03	\\	
\texttt{C207\_C21}	&	1406.55	&	0.005	&	1195.1	&	1211.55	&	13.86	&	262.84	&	\bf1173.7	&	\bf1188.92	&	15.47	&	562.88	\\	
\texttt{C208\_C21}	&	1393.28	&	0.005	&	1193.01	&	1221.4	&	12.34	&	429.15	&	\bf1169.69	&	\bf1188.85	&	14.67	&	652.98	\\	
\texttt{RC101\_C21}	&	2467.62	&	0.004	&	2044.99	&	2274.23	&	7.84	&	294.29	&	\bf1907.52	&	\bf2106.42	&	14.64	&	605.26	\\	
\texttt{RC102\_C21}	&	2385.73	&	0.005	&	2004.78	&	\bf2035.60	&	14.68	&	516.43	&	\bf1834.97	&	2047.79	&	14.16	&	397.12	\\	
\texttt{RC103\_C21}	&	2189.24	&	0.004	&	1747.98	&	1933.49	&	11.68	&	393.23	&	\bf1728.17	&	\bf1846.23	&	15.67	&	470.22	\\	
\texttt{RC104\_C21}	&	1710.42	&	0.004	&	\bf1644.36	&	\bf1686.88	&	1.38	&	322	&	1645.35	&	1688.65	&	1.27	&	372.96	\\	
\texttt{RC105\_C21}	&	2482.3	&	0.005	&	\bf1789.64	&	\bf1821.53	&	26.62	&	471.32	&	1802.85	&	1936.87	&	21.97	&	506.77	\\	
\texttt{RC106\_C21}	&	2142.63	&	0.005	&	1760.23	&	\bf1797.62	&	16.1	&	584.16	&	\bf1750.61	&	1807.42	&	15.64	&	450.61	\\	
\texttt{RC107\_C21}	&	1901.16	&	0.004	&	1687.9	&	\bf1713.75	&	9.86	&	493.35	&	\bf1686.76	&	1719.83	&	9.54	&	681.21	\\	
\texttt{RC108\_C21}	&	1737.71	&	0.005	&	1672.75	&	1676.55	&	3.52	&	440.79	&	\bf1622.76	&	\bf1655.21	&	4.75	&	760.71	\\	
\texttt{RC201\_C21}	&	1809.28	&	0.004	&	\bf1313.01	&	\bf1341.92	&	25.83	&	682.08	&	1318.73	&	1358.75	&	24.9	&	282.6	\\	
\texttt{RC202\_C21}	&	1636.91	&	0.005	&	\bf1218.40	&	1246.29	&	23.86	&	691.5	&	1200.59	&	\bf1230.97	&	24.8	&	531.87	\\	
\texttt{RC203\_C21}	&	1401.03	&	0.005	&	1119.62	&	1140.74	&	18.58	&	589.31	&	\bf1103.43	&	\bf1138.82	&	18.72	&	624.78	\\	
\texttt{RC204\_C21}	&	1267.87	&	0.004	&	1045.72	&	1077.93	&	14.98	&	462.65	&	\bf1040.09	&	\bf1054.96	&	16.79	&	470.14	\\	
\texttt{RC205\_C21}	&	1553.79	&	0.004	&	1223.37	&	1253.27	&	19.34	&	368.42	&	\bf1217.43	&	\bf1245.16	&	19.86	&	356.82	\\	
\texttt{RC206\_C21}	&	1536.28	&	0.004	&	1216.7	&	1235.36	&	19.59	&	495.64	&	\bf1193.11	&	\bf1216.17	&	20.84	&	610.81	\\	
\texttt{RC207\_C21}	&	1424.02	&	0.004	&	1116.3	&	\bf1133.88	&	20.38	&	532.73	&	\bf1106.60	&	1146.08	&	19.52	&	442.03	\\	
\texttt{RC208\_C21}	&	1253.98	&	0.005	&	\bf1038.25	&	1081.38	&	13.76	&	535.7	&	1049.42	&	\bf1067.86	&	14.84	&	516.79	\\	
\texttt{RC101\_C21}	&	2467.62	&	0.004	&	2044.99	&	2274.23	&	7.84	&	294.29	&	\bf1907.52	&	\bf2106.42	&	14.64	&	605.26	\\	
\texttt{RC102\_C21}	&	2385.73	&	0.005	&	2004.78	&	\bf2035.60	&	14.68	&	516.43	&	\bf1834.97	&	2047.79	&	14.16	&	397.12	\\	
\texttt{RC103\_C21}	&	2189.24	&	0.004	&	1747.98	&	1933.49	&	11.68	&	393.23	&	\bf1728.17	&	\bf1846.23	&	15.67	&	470.22	\\	
\texttt{RC104\_C21}	&	1710.42	&	0.004	&	\bf1644.36	&	\bf1686.88	&	1.38	&	322	&	1645.35	&	1688.65	&	1.27	&	372.96	\\	
\texttt{RC105\_C21}	&	2482.3	&	0.005	&	\bf1789.64	&	\bf1821.53	&	26.62	&	471.32	&	1802.85	&	1936.87	&	21.97	&	506.77	\\	
\texttt{RC106\_C21}	&	2142.63	&	0.005	&	1760.23	&	\bf1797.62	&	16.1	&	584.16	&	\bf1750.61	&	1807.42	&	15.64	&	450.61	\\	
\texttt{RC107\_C21}	&	1901.16	&	0.004	&	1687.9	&	\bf1713.75	&	9.86	&	493.35	&	\bf1686.76	&	1719.83	&	9.54	&	681.21	\\	
\texttt{RC108\_C21}	&	1737.71	&	0.005	&	1672.75	&	1676.55	&	3.52	&	440.79	&	\bf1622.76	&	\bf1655.21	&	4.75	&	760.71	\\	
\texttt{RC201\_C21}	&	1809.28	&	0.004	&	\bf1313.01	&	\bf1341.92	&	25.83	&	682.08	&	1318.73	&	1358.75	&	24.9	&	282.6	\\	
\texttt{RC202\_C21}	&	1636.91	&	0.005	&	\bf1218.40	&	1246.29	&	23.86	&	691.5	&	1200.59	&	\bf1230.97	&	24.8	&	531.87	\\	
\texttt{RC203\_C21}	&	1401.03	&	0.005	&	1119.62	&	1140.74	&	18.58	&	589.31	&	\bf1103.43	&	\bf1138.82	&	18.72	&	624.78	\\	
\texttt{RC204\_C21}	&	1267.87	&	0.004	&	1045.72	&	1077.93	&	14.98	&	462.65	&	\bf1040.09	&	\bf1054.96	&	16.79	&	470.14	\\	
\texttt{RC205\_C21}	&	1553.79	&	0.004	&	1223.37	&	1253.27	&	19.34	&	368.42	&	\bf1217.43	&	\bf1245.16	&	19.86	&	356.82	\\	
\texttt{RC206\_C21}	&	1536.28	&	0.004	&	1216.7	&	1235.36	&	19.59	&	495.64	&	\bf1193.11	&	\bf1216.17	&	20.84	&	610.81	\\	
\texttt{RC207\_C21}	&	1424.02	&	0.004	&	1116.3	&	\bf1133.88	&	20.38	&	532.73	&	\bf1106.60	&	1146.08	&	19.52	&	442.03	\\	
\texttt{RC208\_C21}	&	1253.98	&	0.005	&	\bf1038.25	&	1081.38	&	13.76	&	535.7	&	1049.42	&	\bf1067.86	&	14.84	&	516.79	\\	
																						
\hline \hline 																						
\textbf{average}	&	1750.28	&	0.005	&	1433.99	&	1473.82	&	15.8	&	470.97	&	\bf1406.51	&	\bf1460.15	&	16.59	&	537.23	\\	
\hline																						
\end{tabular}}																						
\end{table}

It is also worth noting that the average improvement rate with respect to the solutions of the Clarke-Wright savings heuristic for large clustered problem instances is lower than in the context of the random and random-clustered instances. One reason for this is possibly the assignment of each customer to the nearest satellite in the initial solution construction phase, which provides most probably a better customer-satellite assignment than in the context of random instances.

\subsection{Two Echelon Electric Vehicle Routing Problem with Simultaneous Pickup and Deliveries}

In logistics, simultaneous pickup and delivery (SPD) is a common and crucial operational requirement. It arises when vehicles are tasked with not only delivering goods to customers but also collecting items from customers during the same route. This is particularly relevant in the context of reverse logistics, where returned goods or recyclable materials need to be transported back to depots or processing facilities. Incorporating SPD into vehicle routing models is important for optimizing resource usage and minimizing total transportation costs in real-world logistics operations.

In the 2E-EVRP-SPD variant, each customer can have both a delivery demand $D^{2}_i > 0$ and a pickup demand $P^{2}_i > 0$, which must be serviced by the electric vehicles in the second echelon. These vehicles have a limited capacity, $Q^{2}$, which must accommodate both delivery and pickup items during the route. The main challenge lies in efficiently balancing the load of the vehicles, ensuring that the total load never exceeds the capacity during the tour.

In this section, we present the additional constraints required to handle simultaneous pickup and delivery. The remaining parts of the model follow the base formulation provided in Section \ref{sec:problem}.

\subsubsection{Notation}
\begin{itemize}
    \item $P^{2}_i$: pickup demand for customer $i \in N_C$.
    \item $v^{2}_{ij}$: remaining picked-up cargo on arc $(i,j)$ in the second echelon.
\end{itemize}

\subsubsection{Additional Constraints for Simultaneous Pickup and Deliveries}

The following constraints extend the base 2E-EVRP model~(\ref{m1:1}-\ref{m1:22}) to incorporate simultaneous pickup and delivery operations in the second echelon.

\begin{align}
    & \sum\limits_{i \in N_{DS}, i \neq j}{v^{1}_{ij}}-\sum\limits_{i \in N_{SD}, i \neq j}{v^{1}_{ji}} = P^{1}_{j}, & \forall j \in N_S \label{spd1}
\end{align}
This constraint ensures that the total pickup demand of satellites is met by the large vehicles in the first echelon.

\begin{align}
    & v^{1}_{ij} = 0, & \forall i \in N_D, j \in N_{SD} \label{spd2}
\end{align}
This constraint ensures that the pickup cargo is initialized to zero when large vehicles leave the central warehouses (depots).

\begin{align}
    & u^{1}_{ij} + v^{1}_{ij} \leq Q^{1} * x_{ij}, & \forall i \in N_D, j \in N_{DS}, i \neq j \label{spd3}
\end{align}
This constraint ensures that the combined delivery and pickup load does not exceed the vehicle capacity $Q^{1}$ in the first echelon.

\begin{align}
    & \sum\limits_{l \in N_{C}}{z_{li} * P^{2}_l} = P^{1}_{i}, & \forall i \in N_S \label{spd4}
\end{align}
This constraint ensures that the total pickup demand of the customers assigned to a satellite equals the pickup demand for that satellite.

\begin{align}
    & \sum\limits_{i \in N_{SRC}, i \neq j}{v^{2}_{ji}} - \sum\limits_{i \in N_{RCS}, i \neq j}{v^{2}_{ij}} = P^{2}_{j}, & \forall j \in N_{RC} \label{spd5}
\end{align}
This constraint balances the pickup flow at each customer node in the second echelon, ensuring that the picked-up cargo matches the pickup demand for that customer.

\begin{align}
    & v^{2}_{ij} = 0, & \forall i \in N_S, j \in N_{RCS} \label{spd6}
\end{align}
This constraint ensures that electric vehicles start their route with no picked-up cargo when they leave the satellites.

\begin{align}
    & u^{2}_{ij} + v^{2}_{ij} \leq Q^{2} * y_{ij}, & \forall i \in N_{RCS}, j \in N_{SRC} \label{spd7}
\end{align}
This constraint ensures that the sum of the delivery and pickup loads in the second echelon does not exceed the electric vehicle capacity $Q^{2}$.

These additional constraints allow the model to handle simultaneous pickup and delivery operations in the second echelon, ensuring that pickup demands are met without violating vehicle capacity constraints. The remaining aspects of the model, such as battery constraints and flow conservation, are the same as in the base 2E-EVRP model.

\subsubsection{Computational Results}

We employed the same experimental setup described in the previous analysis. Specifically, CPLEX (version 12.10) and probabilistic implementations of constructive heuristics, including the C\&W Savings and Insertion heuristics from~\cite{akbay2023application}, were applied to all problem instances. CPLEX was given a time limit of 2 hours of CPU time per instance to ensure it could provide optimal or near-optimal solutions wherever feasible. Similarly, the CMSA algorithm (also from~\cite{akbay2023application}) was evaluated under the same constraints, with computation time limits set to 150 CPU seconds for small problem instances and 900 CPU seconds for larger ones. To account for its stochastic behavior, CMSA was executed 10 times on each problem instance, and results were aggregated for analysis. The obtained results are provided in Tables~\ref{c6_tab:results1}--\ref{c6_tab:results5}.

For the smallest problem instances (5 customers) in Table~\ref{c6_tab:results1}, CPLEX achieved optimal solutions for all instances within a minimal average computation time of 5.34 seconds. Both probabilistic constructive heuristics and CMSA matched the quality of solutions provided by CPLEX, yielding optimal solutions for all instances as well. The computation times for these heuristic approaches were exceptionally low (averaging 0.02 seconds for constructive heuristics and 0.01 seconds for CMSA), demonstrating their efficiency. 

For instances with 10 customers in Table~\ref{c6_tab:results2}, CPLEX maintained its ability to find optimal solutions, with an average computation time of 332.58 seconds. The constructive heuristics and CMSA also delivered optimal solutions for all instances, except for minor discrepancies in average solution quality (column 'Avg'). Notably, the CMSA algorithm provided the same best solutions as CPLEX while requiring a significantly shorter average computation time of 0.12 seconds. This highlights CMSA's computational efficiency for slightly larger problems, especially compared to CPLEX's rapidly increasing computation times.

The performance gap between the methods became evident when the problem size increased to 15 customers; see Table~\ref{c6_tab:results3}. CPLEX faced challenges, solving only a subset of instances to optimality and producing an average optimality gap of 11.55\%. The average computation time for CPLEX also increased substantially, reaching an average of 4733.84 seconds, reflecting the algorithm's struggle to handle these larger instances efficiently. The constructive heuristics provided solutions in much shorter computation times (33.71 seconds on average) but demonstrated a drop in solution quality. These heuristics produced higher average distances (column 'Avg') and, in some cases, solutions that significantly deviated from CPLEX's best results. The CMSA algorithm outperformed the constructive heuristics in terms of both solution quality and consistency. CMSA achieved lower average distances and best distances that were comparable to or better than those of the constructive heuristics, with an average computation time of just 11.07 seconds. 

The results also reveal the impact of the simultaneous pickup and delivery constraint on the complexity of the problem. Compared to the base problem (2E-EVRP), the computational time required by CPLEX to find optimal and near-optimal solutions increased significantly, even for small instances. For example, the average time required by CPLEX to solve instances with 5 customers rose from 2.72 seconds in the base problem to 5.34 seconds in this variant. This demonstrates that adding simultaneous pickup and delivery constraints increases the problem's computational complexity, further amplifying the need for efficient heuristic methods like CMSA to handle larger instances effectively.

\begin{table}[h]
\centering
\caption{Computational results for the 2E-EVRP-SPD, small-sized instances with 5 customers.}
\label{c6_tab:results1}
\scalebox{0.75}{
\begin{tabular}{c|ccc|ccc|ccc} \noalign{\hrule height 1pt}
\multicolumn{1}{c|}{\textbf{Instances}} & \multicolumn{3}{c|}{\textbf{CPLEX}} & \multicolumn{3}{c|}{\boldmath{\textbf{CONSTRUCTIVE HEURISTICS}}} & \multicolumn{3}{c}{\boldmath{\textbf{CMSA}}} \\
\textbf{Name} & \textbf{Dist} &  \boldmath{$\overline{t(s)}$}&  \textbf{Gap(\%)}  & \textbf{Dist} & \textbf{Avg} & \boldmath{$\overline{t(s)}$} & \textbf{Dist} & \textbf{Avg} & \boldmath{$\overline{t(s)}$} \\
\hline
\texttt{C101\_C5x} & \bf325.0 & 1.14 & 0.0 & \bf325.0 & 325.0 & 0.0 & \bf325.0 & 325.0 & 0.0 \\
\texttt{C103\_C5x} & \bf298.0 & 0.94 & 0.0 & \bf298.0 & 298.0 & 0.0 & \bf298.0 & 298.0 & 0.0 \\
\texttt{C206\_C5x} & \bf351.0 & 1.13 & 0.0 & \bf351.0 & 351.0 & 0.0 & \bf351.0 & 351.0 & 0.0 \\
\texttt{C208\_C5x} & \bf382.0 & 8.66 & 0.0 & \bf382.0 & 382.0 & 0.16 & \bf382.0 & 382.0 & 0.01 \\
\texttt{R104\_C5x} & \bf316.0 & 1.76 & 0.0 & \bf316.0 & 316.0 & 0.03 & \bf316.0 & 316.0 & 0.0 \\
\texttt{R105\_C5x} & \bf352.0 & 1.36 & 0.0 & 372.0 & 372.0 & 0.0 & \bf352.0 & 352.0 & 0.02 \\
\texttt{R202\_C5x} & \bf348.0 & 5.05 & 0.0 & \bf348.0 & 348.0 & 0.0 & \bf348.0 & 348.0 & 0.01 \\
\texttt{R203\_C5x} & \bf372.0 & 7.84 & 0.0 & \bf372.0 & 372.0 & 0.0 & \bf372.0 & 372.0 & 0.0 \\
\texttt{RC105\_C5x} & \bf356.0 & 18.91 & 0.0 & \bf356.0 & 356.0 & 0.07 & \bf356.0 & 356.0 & 0.01 \\
\texttt{RC108\_C5x} & \bf380.0 & 7.11 & 0.0 & \bf380.0 & 380.0 & 0.0 & \bf380.0 & 380.0 & 0.0 \\
\texttt{RC204\_C5x} & \bf332.0 & 8.39 & 0.0 & 465.0 & 529.0 & 0.0 & \bf332.0 & 332.0 & 0.01 \\
\texttt{RC208\_C5x} & \bf328.0 & 1.8 & 0.0 & \bf328.0 & 328.0 & 0.0 & \bf328.0 & 328.0 & 0.0 \\
\hline \hline
\texttt{average} & \bf345.0 & 5.34 & 0.0 & 357.75 & 363.08 & 0.02 & \bf345.0 & 345.0 & 0.01 \\

\hline
\end{tabular}}
\end{table}

\begin{table}[h]
\centering
\caption{Computational results for the 2E-EVRP-SPD, small-sized instances with 10 customers.}
\label{c6_tab:results2}
\scalebox{0.75}{
\begin{tabular}{c|ccc|ccc|ccc} \noalign{\hrule height 1pt}
\multicolumn{1}{c|}{\textbf{Instances}} & \multicolumn{3}{c|}{\textbf{CPLEX}} & \multicolumn{3}{c|}{\boldmath{\textbf{CONSTRUCTIVE HEURISTICS}}} & \multicolumn{3}{c}{\boldmath{\textbf{CMSA}}} \\
\textbf{Name} & \textbf{Dist} &  \boldmath{$\overline{t(s)}$}&  \textbf{Gap(\%)}  & \textbf{Dist} & \textbf{Avg} & \boldmath{$\overline{t(s)}$} & \textbf{Dist} & \textbf{Avg} & \boldmath{$\overline{t(s)}$} \\
\hline
\texttt{C101\_C10x} & \bf475.0 & 1951.17 & 0.0 & \bf475.0 & 475.0 & 2.29 & \bf475.0 & 475.0 & 0.06 \\
\texttt{C104\_C10x} & \bf413.0 & 33.61 & 0.0 & \bf413.0 & 416.5 & 17.49 & \bf413.0 & 413.0 & 0.38 \\
\texttt{C202\_C10x} & \bf369.0 & 28.38 & 0.0 & \bf369.0 & 369.0 & 0.29 & \bf369.0 & 369.0 & 0.06 \\
\texttt{C205\_C10x} & \bf402.0 & 169.0 & 0.0 & \bf402.0 & 402.6 & 2.74 & \bf402.0 & 402.0 & 0.1 \\
\texttt{R102\_C10x} & \bf359.0 & 29.05 & 0.0 & \bf359.0 & 359.0 & 0.99 & \bf359.0 & 359.0 & 0.13 \\
\texttt{R103\_C10x} & \bf330.0 & 81.94 & 0.0 & \bf330.0 & 330.0 & 0.69 & \bf330.0 & 330.0 & 0.06 \\
\texttt{R201\_C10x} & \bf349.0 & 62.05 & 0.0 & \bf349.0 & 349.0 & 0.1 & \bf349.0 & 349.0 & 0.05 \\
\texttt{R203\_C10x} & \bf436.0 & 772.11 & 0.0 & \bf436.0 & 436.0 & 19.78 & \bf436.0 & 436.0 & 0.12 \\
\texttt{RC102\_C10x} & \bf455.0 & 59.04 & 0.0 & \bf455.0 & 455.0 & 2.05 & \bf455.0 & 455.0 & 0.12 \\
\texttt{RC108\_C10x} & \bf472.0 & 369.0 & 0.0 & \bf472.0 & 473.8 & 2.41 & \bf472.0 & 472.0 & 0.11 \\
\texttt{RC201\_C10x} & \bf395.0 & 118.62 & 0.0 & \bf395.0 & 395.0 & 0.63 & \bf395.0 & 395.0 & 0.15 \\
\texttt{RC205\_C10x} & \bf487.0 & 317.04 & 0.0 & \bf487.0 & 487.5 & 21.32 & \bf487.0 & 487.0 & 0.13 \\
\hline \hline
\texttt{average} & \bf411.83 & 332.58 & 0.0 & \bf411.83 & 412.37 & 5.9 & \bf411.83 & 411.83 & 0.12 \\

\hline
\end{tabular}}
\end{table}

\begin{table}[h]
\centering
\caption{Computational results for the 2E-EVRP-SPD, small-sized instances with 15 customers.}
\label{c6_tab:results3}
\scalebox{0.75}{
\begin{tabular}{c|ccc|ccc|ccc} \noalign{\hrule height 1pt}
\multicolumn{1}{c|}{\textbf{Instances}} & \multicolumn{3}{c|}{\textbf{CPLEX}} & \multicolumn{3}{c|}{\boldmath{\textbf{CONSTRUCTIVE HEURISTICS}}} & \multicolumn{3}{c}{\boldmath{\textbf{CMSA}}} \\
\textbf{Name} & \textbf{Dist} &  \boldmath{$\overline{t(s)}$}&  \textbf{Gap(\%)}  & \textbf{Dist} & \textbf{Avg} & \boldmath{$\overline{t(s)}$} & \textbf{Dist} & \textbf{Avg} & \boldmath{$\overline{t(s)}$} \\
\hline
\texttt{C103\_C15x} & \bf450.0 & 5743.48 & 12.66 & 450.0 & 467.7 & 66.5 & \bf450.0 & 450.0 & 2.47 \\
\texttt{C106\_C15x} & \bf407.0 & 3590.9 & 9.24 & 415.0 & 432.0 & 22.82 & \bf407.0 & 407.0 & 2.26 \\
\texttt{C202\_C15x} & \bf460.0 & 7034.28 & 4.32 & \bf460.0 & 460.9 & 46.35 & \bf460.0 & 460.0 & 0.56 \\
\texttt{C208\_C15x} & \bf449.0 & 3591.46 & 10.49 & \bf449.0 & 449.0 & 7.59 & \bf449.0 & 449.0 & 0.41 \\
\texttt{R102\_C15x} & 466.0 & 7050.44 & 17.49 & 496.0 & 499.0 & 43.38 & \bf457.0 & 457.0 & 15.92 \\
\texttt{R105\_C15x} & 514.0 & 4729.45 & 27.34 & 462.0 & 480.3 & 17.99 & \bf449.0 & 456.9 & 71.46 \\
\texttt{R202\_C15x} & \bf429.0 & 3590.65 & 4.89 & \bf29.0 & 429.0 & 13.11 & \bf429.0 & 429.0 & 0.77 \\
\texttt{R209\_C15x} & \bf409.0 & 3591.36 & 0.48 & \bf409.0 & 409.5 & 17.28 & \bf409.0 & 409.0 & 0.43 \\
\texttt{RC103\_C15x} & \bf469.0 & 3590.66 & 15.14 & 471.0 & 494.8 & 84.42 & \bf469.0 & 469.0 & 3.57 \\
\texttt{RC108\_C15x} & 535.0 & 7110.32 & 18.68 & 563.0 & 570.2 & 43.57 & \bf534.0 & 534.0 & 33.74 \\
\texttt{RC202\_C15x} & 455.0 & 3591.53 & 6.59 & \bf454.0 & 454.6 & 10.46 & \bf454.0 & 454.0 & 0.62 \\
\texttt{RC204\_C15x} & \bf435.0 & 3591.55 & 11.28 & \bf435.0 & 438.6 & 31.07 & \bf435.0 & 435.0 & 0.66 \\
\hline \hline
\texttt{average} & 456.5 & 4733.84 & 11.55 & 457.75 & 465.47 & 33.71 & \bf450.17 & 450.82 & 11.07 \\

\hline
\end{tabular}}
\end{table}

Across all large-sized instances, CMSA consistently outperformed the constructive heuristics in terms of solution quality. On average, the best solutions obtained by CMSA were 13.99\% better than those generated by constructive heuristics. Constructive heuristics required more CPU time on average (713.47 seconds) compared to CMSA (486.45 seconds), making CMSA not only superior in terms of solution quality but also more computationally efficient.

\begin{table}[h]
\centering
\caption{Computational results for the 2E-EVRP-SPD, large-sized instances with 100 customers.}
\label{c6_tab:results5}
\scalebox{0.75}{
\begin{tabular}{c|ccc|ccc} \noalign{\hrule height 1pt}
\multicolumn{1}{c|}{\textbf{Instances}} & \multicolumn{3}{c|}{\boldmath{\textbf{CONSTRUCTIVE HEURISTICS}}} & \multicolumn{3}{c}{\boldmath{\textbf{CMSA}}} \\
\textbf{Name}  & \textbf{Dist} & \textbf{Avg} & \boldmath{$\overline{t(s)}$} & \textbf{Dist} & \textbf{Avg} & \boldmath{$\overline{t(s)}$} \\
\hline
\texttt{C101\_21x} & 1270.0 & 1345.7 & 676.07 & \bf1151.0 & 1220.1 & 578.39 \\
\texttt{C102\_21x} & 1201.0 & 1211.8 & 822.66 & \bf1128.0 & 1207.4 & 646.38 \\
\texttt{C103\_21x} & 1172.0 & 1304.6 & 548.79 & \bf1141.0 & 1229.0 & 527.52 \\
\texttt{C104\_21x} & 1201.0 & 1291.1 & 577.71 & \bf1135.0 & 1212.1 & 518.02 \\
\texttt{C105\_21x} & 1201.0 & 1267.4 & 607.71 & \bf1153.0 & 1243.2 & 588.66 \\
\texttt{C106\_21x} & 1258.0 & 1334.7 & 704.1 & \bf1168.0 & 1263.0 & 709.9 \\
\texttt{C107\_21x} & 1234.0 & 1327.4 & 768.02 & \bf1152.0 & 1231.4 & 601.34 \\
\texttt{C108\_21x} & 1198.0 & 1287.8 & 730.27 & \bf1157.0 & 1250.3 & 493.21 \\
\texttt{C109\_21x} & 1164.0 & 1246.6 & 697.89 & \bf1160.0 & 1248.1 & 643.57 \\
\texttt{C201\_21x} & 1100.0 & 1204.1 & 825.08 & \bf926.0 & 965.7 & 212.51 \\
\texttt{C202\_21x} & 1114.0 & 1172.5 & 718.43 & \bf921.0 & 936.5 & 71.49 \\
\texttt{C203\_21x} & 1132.0 & 1159.4 & 747.78 & \bf924.0 & 944.9 & 130.25 \\
\texttt{C204\_21x} & 1128.0 & 1137.0 & 711.06 & \bf927.0 & 941.0 & 172.41 \\
\texttt{C205\_21x} & 1005.0 & 1170.4 & 707.76 & \bf930.0 & 965.3 & 262.22 \\
\texttt{C206\_21x} & 1094.0 & 1210.2 & 832.48 & \bf930.0 & 944.3 & 144.6 \\
\texttt{C207\_21x} & 1184.0 & 1272.7 & 746.98 & \bf929.0 & 940.3 & 140.4 \\
\texttt{C208\_21x} & 1179.0 & 1217.1 & 800.79 & \bf920.0 & 933.4 & 159.99 \\
\texttt{R101\_21x} & 1406.0 & 1491.0 & 768.4 & 1\bf220.0 & 1256.2 & 569.13 \\
\texttt{R102\_21x} & 1433.0 & 1518.9 & 760.51 & \bf1211.0 & 1250.7 & 503.14 \\
\texttt{R103\_21x} & 1326.0 & 1463.6 & 792.14 & \bf1230.0 & 1237.1 & 317.89 \\
\texttt{R104\_21x} & 1340.0 & 1516.2 & 691.23 & \bf1222.0 & 1250.8 & 600.79 \\
\texttt{R105\_21x} & 1308.0 & 1416.5 & 681.63 & \bf1211.0 & 1251.0 & 469.33 \\
\texttt{R106\_21x} & 1453.0 & 1559.1 & 712.15 & \bf1217.0 & 1244.8 & 623.15 \\
\texttt{R107\_21x} & 1448.0 & 1551.7 & 429.13 & \bf1193.0 & 1238.5 & 597.11 \\
\texttt{R108\_21x} & 1509.0 & 1538.4 & 797.79 & \bf1213.0 & 1237.8 & 677.09 \\
\texttt{R109\_21x} & 1456.0 & 1520.2 & 420.15 & \bf1202.0 & 1241.0 & 506.06 \\
\texttt{R110\_21x} & 1411.0 & 1472.2 & 549.7 & \bf1224.0 & 1246.7 & 662.1 \\
\texttt{R111\_21x} & 1381.0 & 1467.9 & 617.77 & \bf1221.0 & 1268.0 & 707.78 \\
\texttt{R112\_21x} & 1387.0 & 1455.7 & 716.8 & \bf1226.0 & 1251.4 & 758.57 \\
\texttt{R201\_21x} & 1082.0 & 1228.7 & 712.14 & \bf911.0 & 927.5 & 426.45 \\
\texttt{R202\_21x} & 1141.0 & 1204.2 & 790.1 & \bf905.0 & 931.8 & 359.69 \\
\texttt{R203\_21x} & 1066.0 & 1193.5 & 659.67 & \bf903.0 & 936.4 & 297.47 \\
\texttt{R204\_21x} & 1115.0 & 1232.7 & 568.3 & \bf905.0 & 932.0 & 364.08 \\
\texttt{R205\_21x} & 1116.0 & 1263.5 & 543.53 & \bf911.0 & 930.6 & 555.93 \\
\texttt{R206\_21x} & 1131.0 & 1276.3 & 750.39 & \bf915.0 & 928.7 & 365.3 \\
\texttt{R207\_21x} & 1302.0 & 1374.4 & 641.82 & \bf909.0 & 933.4 & 436.56 \\
\texttt{R208\_21x} & 1093.0 & 1186.0 & 773.05 & \bf905.0 & 936.2 & 266.31 \\
\texttt{R209\_21x} & 1148.0 & 1194.9 & 779.06 & \bf923.0 & 938.8 & 483.18 \\
\texttt{R210\_21x} & 1138.0 & 1203.8 & 547.65 & \bf920.0 & 943.2 & 183.56 \\
\texttt{R211\_21x} & 1068.0 & 1246.1 & 728.98 & \bf905.0 & 928.2 & 436.82 \\
\texttt{RC101\_21x} & 1656.0 & 1744.6 & 861.35 & \bf1347.0 & 1430.5 & 642.71 \\
\texttt{RC102\_21x} & 1569.0 & 1598.8 & 686.78 & \bf1341.0 & 1414.2 & 637.17 \\
\texttt{RC103\_21x} & 1540.0 & 1663.4 & 713.22 & \bf1344.0 & 1398.6 & 652.49 \\
\texttt{RC104\_21x} & 1487.0 & 1649.0 & 830.27 & \bf1364.0 & 1465.6 & 732.14 \\
\texttt{RC105\_21x} & 1494.0 & 1612.1 & 704.97 & \bf1364.0 & 1435.7 & 536.79 \\
\texttt{RC106\_21x} & 1481.0 & 1565.1 & 701.24 & \bf1380.0 & 1451.8 & 827.09 \\
\texttt{RC107\_21x} & 1554.0 & 1602.6 & 712.99 & \bf1383.0 & 1424.9 & 723.35 \\
\texttt{RC108\_21x} & 1613.0 & 1717.4 & 808.18 & \bf1343.0 & 1410.4 & 661.61 \\
\texttt{RC201\_21x} & 1145.0 & 1207.7 & 797.44 & \bf931.0 & 964.4 & 390.73 \\
\texttt{RC202\_21x} & 1096.0 & 1183.8 & 854.73 & \bf918.0 & 936.7 & 534.27 \\
\texttt{RC203\_21x} & 1085.0 & 1209.5 & 758.38 & \bf919.0 & 942.2 & 610.38 \\
\texttt{RC204\_21x} & 1073.0 & 1302.4 & 778.89 & \bf921.0 & 945.6 & 728.77 \\
\texttt{RC205\_21x} & 1180.0 & 1249.4 & 694.42 & \bf919.0 & 936.6 & 451.25 \\
\texttt{RC206\_21x} & 1113.0 & 1210.4 & 753.26 & \bf921.0 & 938.6 & 387.53 \\
\texttt{RC207\_21x} & 1037.0 & 1327.9 & 788.15 & \bf919.0 & 949.1 & 498.33 \\
\texttt{RC208\_21x} & 1245.0 & 1315.4 & 854.11 & \bf924.0 & 947.4 & 458.41 \\
\hline \hline
\texttt{average} & 1258.23 & 1355.28 & 713.47 & \bf1082.0 & 1123.38 & 486.45 \\

\hline
\end{tabular}}
\end{table}

\subsection{Two Echelon Electric Vehicle Routing Problem with Partial Deliveries}

In practical logistics scenarios, high-demand customers or those in geographically challenging locations often benefit from being served by multiple vehicles. The 2E-EVRP with Partial Deliveries (2E-EVRP-PD) introduces the flexibility to split delivery demands across two vehicles, enabling more efficient and adaptable routing strategies. This variant aligns closely with real-world logistics requirements, where constraints such as vehicle capacity, time windows, or geographical barriers necessitate such an approach.

\subsubsection{Modeling Partial Deliveries Without Structural Changes}

Unlike traditional problem formulations that restrict each customer to a single vehicle visit, the 2E-EVRP-PD permits two vehicles to share the delivery task for a single customer by splitting the total demand into partial loads. This adaptation is implemented at the dataset level, avoiding the need to modify the mathematical constraints of the underlying optimization model. Instead, the flexibility is integrated into the dataset itself, which simplifies the implementation while retaining the original problem structure.

\subsubsection{Dataset Adjustments for Partial Deliveries}

The dataset was modified to accommodate partial deliveries as follows:

\begin{enumerate}
    \item \textbf{Duplication of Demand Points}: Each customer \(c_i\) in the original dataset is duplicated to create two new demand points, \(c_i^{(1)}\) and \(c_i^{(2)}\). These duplicates represent separate delivery targets for the two vehicles that will meet the customer’s demand.

    \item \textbf{Division of Demand}: The original demand \(q_i\) for each customer is divided between \(c_i^{(1)}\) and \(c_i^{(2)}\). For simplicity and consistency, the demand is split equally between the two points:
    \[
    q_i^{(1)} = q_i^{(2)} = \frac{q_i}{2}.
    \]

    \item \textbf{Optional Division Rate Parameter}: To add further flexibility and realism, an optional parameter, called the \textit{division rate}, can be introduced. This parameter specifies the proportion of the original demand \(q_i\) that can be allocated to one vehicle, with the remainder served by another. The division rate can vary between 20\% and 50\%, allowing for asymmetric splitting:
    \[
    q_i^{(1)} = q_i \times \frac{n}{100}, \quad q_i^{(2)} = q_i \times \left(1 - \frac{n}{100}\right).
    \]
    Here, \(n\%\) is the division rate assigned randomly or based on specific customer needs.
\end{enumerate}

\subsubsection{Computational Results}

In this section, we applied the same experimental setup and the same algorithms as in the base problem (2E-EVRP) described in Section~\ref{sec:base-problem}. The obtained results are provided in Tables~\ref{c6_tab:pd-results1}--\ref{c6_tab:pd-results5}.

For the smallest problem instances with 5 customers (see Table~\ref{c6_tab:pd-results1}), CPLEX consistently provided optimal solutions for all instances, requiring an average computation time of 66.41 seconds, which is significantly higher compared to the base 2E-EVRP and the 2E-EVRP-SPD. This indicates the additional complexity introduced by the partial delivery scenario as the problem size is effectively doubled, with each customer duplicated and assigned partial demands. Both probabilistic constructive heuristics and CMSA matched the quality of CPLEX solutions, achieving optimal results with negligible computation times. Constructive heuristics required an average of 1.74 seconds, while CMSA was even faster, with an average computation time of 0.28 seconds. This demonstrates the efficiency of heuristic methods in handling such constraints, even for small problem sizes.

For instances with 10 customers (see Table~\ref{c6_tab:pd-results2}), CPLEX struggled to maintain optimal performance, achieving an average optimality gap of 8.23\%. The computation time surged to 3227.61 seconds on average, reflecting the increased problem complexity due to partial deliveries. In contrast, CMSA consistently delivered solutions of comparable quality to those provided by CPLEX, with an average computation time of just 0.85 seconds. The probabilistic constructive heuristics also performed well, with a slight drop in solution quality but maintaining low computational overhead (0.49 seconds on average). CMSA's ability to achieve high-quality solutions in significantly shorter times further underscores its suitability for addressing the challenges posed by the partial delivery constraint.

As the problem size increased to 15 customers (see Table~\ref{c6_tab:pd-results3}), the gap between CPLEX and the heuristic methods became more apparent. CPLEX faced significant difficulties, producing an average optimality gap of 29.34\% and requiring 5592.91 seconds on average. The constructive heuristics and CMSA, on the other hand, maintained their efficiency, with CMSA outperforming constructive heuristics in both solution quality and computation time. CMSA achieved an average computation time of 12.76 seconds while delivering solutions closer to optimal, demonstrating its robustness for medium-sized problems with complex constraints.

\begin{table}[h]
\centering
\caption{Computational results for the 2E-EVRP-PD, small-sized instances with 5 customers.}
\label{c6_tab:pd-results1}
\scalebox{0.75}{
\begin{tabular}{c|ccc|ccc|ccc} \noalign{\hrule height 1pt}
\multicolumn{1}{c|}{\textbf{Instances}} & \multicolumn{3}{c|}{\textbf{CPLEX}} & \multicolumn{3}{c|}{\boldmath{\textbf{CONSTRUCTIVE HEURISTICS}}} & \multicolumn{3}{c}{\boldmath{\textbf{CMSA}}} \\
\textbf{Name} & \textbf{Dist} &  \boldmath{$\overline{t(s)}$}&  \textbf{Gap(\%)}  & \textbf{Dist} & \textbf{Avg} & \boldmath{$\overline{t(s)}$} & \textbf{Dist} & \textbf{Avg} & \boldmath{$\overline{t(s)}$} \\
\hline
\texttt{C101\_C5x} & 333.0 & 6.68 & 0.0 & \bf257.0 & 257.0 & 0.0 & \bf257.0 & 257.0 & 0.0 \\
\texttt{C103\_C5x} & 298.0 & 4.6 & 0.0 & \bf269.0 & 269.0 & 0.0 & \bf269.0 & 269.0 & 0.0 \\
\texttt{C206\_C5x} & 351.0 & 49.15 & 0.0 & \bf333.0 & 333.0 & 0.0 & \bf333.0 & 333.0 & 0.0 \\
\texttt{C208\_C5x} & \bf326.0 & 108.79 & 0.0 & 365.0 & 365.0 & 0.02 & 365.0 & 365.0 & 0.01 \\
\texttt{R104\_C5x} & 329.0 & 95.24 & 0.0 & \bf277.0 & 277.0 & 0.0 & \bf277.0 & 277.0 & 0.0 \\
\texttt{R105\_C5x} & \bf352.0 & 33.22 & 0.0 & \bf352.0 & 352.0 & 0.69 & \bf352.0 & 352.0 & 1.0 \\
\texttt{R202\_C5x} & 330.0 & 63.43 & 0.0 & \bf312.0 & 312.0 & 0.0 & \bf312.0 & 312.0 & 0.02 \\
\texttt{R203\_C5x} & 372.0 & 84.19 & 0.0 & \bf309.0 & 309.6 & 0.02 & \bf309.0 & 309.2 & 0.06 \\
\texttt{RC105\_C5x} & \bf359.0 & 94.12 & 0.0 & 386.0 & 386.0 & 0.0 & 386.0 & 386.0 & 0.0 \\
\texttt{RC108\_C5x} & 380.0 & 120.36 & 0.0 & \bf369.0 & 369.0 & 0.2 & \bf369.0 & 369.0 & 0.01 \\
\texttt{RC204\_C5x} & 332.0 & 111.98 & 0.0 & \bf293.0 & 293.1 & 19.82 & \bf293.0 & 293.3 & 2.23 \\
\texttt{RC208\_C5x} & 328.0 & 25.22 & 0.0 & \bf307.0 & 307.0 & 0.13 & \bf307.0 & 307.0 & 0.03 \\
\hline \hline
\texttt{average} & 340.83 & 66.41 & 0.0 & 319.08 & 319.14 & 1.74 & \bf319.08 & 319.12 & 0.28 \\

\hline
\end{tabular}}
\end{table}

\begin{table}[h]
\centering
\caption{Computational results for the 2E-EVRP-PD, small-sized instances with 10 customers.}
\label{c6_tab:pd-results2}
\scalebox{0.75}{
\begin{tabular}{c|ccc|ccc|ccc} \noalign{\hrule height 1pt}
\multicolumn{1}{c|}{\textbf{Instances}} & \multicolumn{3}{c|}{\textbf{CPLEX}} & \multicolumn{3}{c|}{\boldmath{\textbf{CONSTRUCTIVE HEURISTICS}}} & \multicolumn{3}{c}{\boldmath{\textbf{CMSA}}} \\
\textbf{Name} & \textbf{Dist} &  \boldmath{$\overline{t(s)}$}&  \textbf{Gap(\%)}  & \textbf{Dist} & \textbf{Avg} & \boldmath{$\overline{t(s)}$} & \textbf{Dist} & \textbf{Avg} & \boldmath{$\overline{t(s)}$} \\
\hline
\texttt{C101\_C10x} & 490.0 & 3591.43 & 14.93 & \bf346.0 & 346.1 & 0.34 & \bf346.0 & 346.2 & 0.98 \\
\texttt{C104\_C10x} & 523.0 & 3591.48 & 15.31 & \bf389.0 & 389.0 & 0.01 & \bf389.0 & 389.0 & 1.02 \\
\texttt{C202\_C10x} & 395.0 & 3591.47 & 6.57 & \bf380.0 & 380.0 & 0.0 & \bf380.0 & 380.0 & 0.01 \\
\texttt{C205\_C10x} & 402.0 & 3591.45 & 6.63 & \bf345.0 & 345.4 & 0.13 & \bf345.0 & 346.2 & 0.0 \\
\texttt{R102\_C10x} & \bf406.0 & 3394.96 & 0.0 & 487.0 & 503.3 & 0.66 & 498.0 & 499.5 & 6.84 \\
\texttt{R103\_C10x} & 425.0 & 3591.28 & 9.06 & \bf298.0 & 298.0 & 4.39 & \bf298.0 & 298.0 & 0.21 \\
\texttt{R201\_C10x} & 378.0 & 537.2 & 0.0 & 290.0 & \bf290.6 & 0.3 & \bf290.0 & 291.2 & 0.22 \\
\texttt{R203\_C10x} & 428.0 & 3591.36 & 18.83 & \bf361.0 & 361.0 & 0.01 & \bf361.0 & 361.0 & 0.12 \\
\texttt{RC102\_C10x} & 513.0 & 3591.39 & 6.03 & \bf357.0 & 357.0 & 0.0 & \bf357.0 & 357.0 & 0.01 \\
\texttt{RC108\_C10x} & 539.0 & 3591.4 & 9.9 & \bf439.0 & 439.0 & 0.01 & \bf439.0 & 439.0 & 0.0 \\
\texttt{RC201\_C10x} & 395.0 & 2476.54 & 0.0 & \bf328.0 & 328.0 & 0.01 & \bf328.0 & 328.0 & 0.01 \\
\texttt{RC205\_C10x} & 486.0 & 3591.39 & 11.5 & \bf329.0 & 329.0 & 0.01 & \bf329.0 & 329.0 & 0.73 \\
\hline \hline
\texttt{average} & 448.33 & 3227.61 & 8.23 & 362.42 & 363.87 & 0.49 & 363.33 & 363.68 & 0.85 \\

\hline
\end{tabular}}
\end{table}

\begin{table}[h]
\centering
\caption{Computational results for the 2E-EVRP-PD, small-sized instances with 15 customers.}
\label{c6_tab:pd-results3}
\scalebox{0.75}{
\begin{tabular}{c|ccc|ccc|ccc} \noalign{\hrule height 1pt}
\multicolumn{1}{c|}{\textbf{Instances}} & \multicolumn{3}{c|}{\textbf{CPLEX}} & \multicolumn{3}{c|}{\boldmath{\textbf{CONSTRUCTIVE HEURISTICS}}} & \multicolumn{3}{c}{\boldmath{\textbf{CMSA}}} \\
\textbf{Name} & \textbf{Dist} &  \boldmath{$\overline{t(s)}$}&  \textbf{Gap(\%)}  & \textbf{Dist} & \textbf{Avg} & \boldmath{$\overline{t(s)}$} & \textbf{Dist} & \textbf{Avg} & \boldmath{$\overline{t(s)}$} \\
\hline
\texttt{C103\_C15x} & 501.0 & 3591.49 & 31.08 & \bf411.0 & 425.4 & 3.85 & \bf411.0 & 418.4 & 4.09 \\
\texttt{C106\_C15x} & 511.0 & 6901.7 & 32.62 & \bf421.0 & 421.0 & 0.64 & \bf421.0 & 421.0 & 3.27 \\
\texttt{C202\_C15x} & 522.0 & 6928.13 & 30.94 & 375.0 & 420.6 & 0.02 & \bf359.0 & 406.4 & 5.55 \\
\texttt{C208\_C15x} & \bf502.0 & 4016.58 & 27.1 & 559.0 & 559.2 & 0.01 & 546.0 & 557.0 & 2.84 \\
\texttt{R102\_C15x} & 585.0 & 7020.55 & 40.12 & 523.0 & 527.8 & 32.94 & \bf520.0 & 526.5 & 23.88 \\
\texttt{R105\_C15x} & 569.0 & 3591.25 & 37.38 & \bf468.0 & 468.0 & 1.63 & \bf468.0 & 468.3 & 3.77 \\
\texttt{R202\_C15x} & 480.0 & 4531.11 & 29.15 & 410.0 & 421.4 & 23.1 & \bf376.0 & 409.4 & 30.48 \\
\texttt{R209\_C15x} & 444.0 & 6922.0 & 19.32 & 353.0 & 354.7 & 25.31 & \bf339.0 & 356.5 & 22.15 \\
\texttt{RC103\_C15x} & \bf507.0 & 6879.0 & 25.51 & 536.0 & 536.0 & 6.54 & 536.0 & 536.0 & 0.99 \\
\texttt{RC108\_C15x} & 562.0 & 7162.37 & 32.03 & \bf497.0 & 500.8 & 0.26 & \bf497.0 & 498.9 & 0.7 \\
\texttt{RC202\_C15x} & 449.0 & 3591.55 & 26.48 & 424.0 & 427.1 & 62.36 & \bf416.0 & 433.8 & 35.57 \\
\texttt{RC204\_C15x} & 447.0 & 5979.15 & 20.32 & 465.0 & 480.3 & 2.49 & \bf433.0 & 487.5 & 19.86 \\
\hline \hline
\texttt{average} & 506.58 & 5592.91 & 29.34 & 453.5 & 461.86 & 13.26 & \bf443.5 & 459.98 & 12.76 \\

\hline
\end{tabular}}
\end{table}

For large problem instances with 100 customers (see Table~\ref{c6_tab:pd-results5}), CPLEX was excluded from comparisons due to the impracticality of solving these instances within the given time constraints. The focus was on comparing the probabilistic constructive heuristics with CMSA. While CMSA generally achieved better performance in terms of average solution quality across all instances, there were specific instances where the constructive heuristics produced better individual solutions. The computational time for CMSA remained competitive, averaging 250.86 seconds compared to 293.88 seconds for constructive heuristics. However, the use of an exact solver in CMSA may have contributed to slightly longer runtimes in certain cases. Overall, CMSA demonstrated robustness and superior performance in delivering consistently high-quality solutions, particularly when averaged across all instances.

\begin{table}[h]
\centering
\caption{Computational results for the 2E-EVRP-PD, large-sized instances with 100 customers.}
\label{c6_tab:pd-results5}
\scalebox{0.75}{
\begin{tabular}{c|ccc|ccc} \noalign{\hrule height 1pt}
\multicolumn{1}{c|}{\textbf{Instances}} & \multicolumn{3}{c|}{\boldmath{\textbf{CONSTRUCTIVE HEURISTICS}}} & \multicolumn{3}{c}{\boldmath{\textbf{CMSA}}} \\
\textbf{Name}  & \textbf{Dist} & \textbf{Avg} & \boldmath{$\overline{t(s)}$} & \textbf{Dist} & \textbf{Avg} & \boldmath{$\overline{t(s)}$} \\
\hline
\texttt{C101\_21x} & \bf1174.0 & 1181.2 & 238.38 & 1177.0 & 1189.3 & 240.78 \\
\texttt{C102\_21x} & 1174.0 & 1189.8 & 276.31 & \bf1165.0 & 1195.9 & 342.83 \\
\texttt{C103\_21x} & 1176.0 & 1184.6 & 366.67 & \bf1169.0 & 1192.7 & 137.41 \\
\texttt{C104\_21x} & 1176.0 & 1201.8 & 153.16 & \bf1170.0 & 1191.5 & 291.6 \\
\texttt{C105\_21x} & \bf1170.0 & 1175.6 & 213.97 & 1185.0 & 1205.0 & 389.94 \\
\texttt{C106\_21x} & \bf1163.0 & 1176.5 & 482.86 & 1176.0 & 1193.4 & 162.82 \\
\texttt{C107\_21x} & 1177.0 & 1180.2 & 53.88 & \bf1163.0 & 1189.0 & 139.34 \\
\texttt{C108\_21x} & \bf1167.0 & 1191.5 & 326.62 & 1171.0 & 1187.7 & 199.76 \\
\texttt{C109\_21x} & 1178.0 & 1228.2 & 54.67 & \bf1175.0 & 1195.5 & 256.58 \\
\texttt{C201\_21x} & \bf996.0 & 1002.9 & 65.56 & 994.0 & 1024.9 & 303.23 \\
\texttt{C202\_21x} & 993.0 & 994.2 & 786.8 & \bf983.0 & 1007.9 & 158.09 \\
\texttt{C203\_21x} & 986.0 & 992.6 & 325.23 & \bf979.0 & 1002.6 & 210.93 \\
\texttt{C204\_21x} & 1005.0 & 1007.5 & 85.96 & \bf980.0 & 1015.7 & 225.0 \\
\texttt{C205\_21x} & 999.0 & 1013.3 & 671.89 & \bf997.0 & 1013.9 & 140.67 \\
\texttt{C206\_21x} & \bf964.0 & 977.9 & 336.87 & 982.0 & 1022.5 & 245.72 \\
\texttt{C207\_21x} & 1000.0 & 1025.3 & 88.24 & \bf977.0 & 1024.2 & 396.1 \\
\texttt{C208\_21x} & \bf954.0 & 1004.1 & 176.53 & 971.0 & 998.8 & 310.93 \\
\texttt{R101\_21x} & \bf1431.0 & 1566.6 & 223.27 & 1407.0 & 1442.9 & 87.28 \\
\texttt{R102\_21x} & 1411.0 & 1455.4 & 219.52 & \bf1396.0 & 1482.1 & 296.93 \\
\texttt{R103\_21x} & \bf1356.0 & 1450.9 & 444.65 & 1418.0 & 1463.9 & 198.14 \\
\texttt{R104\_21x} & \bf1390.0 & 1458.9 & 111.17 & 1415.0 & 1464.1 & 225.1 \\
\texttt{R105\_21x} & 1448.0 & 1458.4 & 110.71 & \bf1438.0 & 1457.4 & 104.82 \\
\texttt{R106\_21x} & \bf1378.0 & 1446.1 & 67.63 & 1448.0 & 1474.0 & 90.37 \\
\texttt{R107\_21x} & 1400.0 & 1477.8 & 213.09 & \bf1375.0 & 1444.5 & 256.25 \\
\texttt{R108\_21x} & 1451.0 & 1467.7 & 290.06 & \bf1390.0 & 1469.1 & 103.0 \\
\texttt{R109\_21x} & 1428.0 & 1617.9 & 3.05 & \bf1384.0 & 1456.9 & 119.92 \\
\texttt{R110\_21x} & 1412.0 & 1422.6 & 29.3 & \bf1404.0 & 1477.0 & 41.18 \\
\texttt{R111\_21x} & 1403.0 & 1486.9 & 104.29 & \bf1382.0 & 1472.9 & 161.01 \\
\texttt{R112\_21x} & 1433.0 & 1455.7 & 10.28 & \bf1408.0 & 1445.4 & 150.75 \\
\texttt{R201\_21x} & \bf910.0 & 978.5 & 196.48 & 927.0 & 1003.4 & 355.14 \\
\texttt{R202\_21x} & 937.0 & 989.2 & 254.19 & \bf935.0 & 1012.7 & 283.4 \\
\texttt{R203\_21x} & 967.0 & 1024.3 & 459.05 & \bf924.0 & 1004.2 & 283.96 \\
\texttt{R204\_21x} & \bf916.0 & 1011.9 & 270.18 & \bf916.0 & 981.1 & 199.82 \\
\texttt{R205\_21x} & 972.0 & 1004.5 & 226.5 & \bf909.0 & 986.3 & 235.75 \\
\texttt{R206\_21x} & 972.0 & 989.6 & 623.94 & \bf948.0 & 1014.4 & 299.64 \\
\texttt{R207\_21x} & \bf972.0 & 985.6 & 141.41 & \bf927.0 & 1018.6 & 354.88 \\
\texttt{R208\_21x} & 930.0 & 980.0 & 254.5 & \bf905.0 & 984.9 & 407.84 \\
\texttt{R209\_21x} & 973.0 & 990.0 & 483.75 & \bf945.0 & 1003.0 & 369.62 \\
\texttt{R210\_21x} & \bf912.0 & 985.6 & 578.24 & 966.0 & 1007.5 & 361.27 \\
\texttt{R211\_21x} & \bf913.0 & 974.6 & 392.18 & 919.0 & 996.9 & 310.07 \\
\texttt{RC101\_21x} & \bf1504.0 & 1505.8 & 460.43 & 1521.0 & 1575.5 & 255.9 \\
\texttt{RC102\_21x} & 1509.0 & 1581.1 & 369.87 & \bf1433.0 & 1528.1 & 197.18 \\
\texttt{RC103\_21x} & 1493.0 & 1553.6 & 297.44 & \bf1480.0 & 1546.0 & 243.19 \\
\texttt{RC104\_21x} & 1508.0 & 1587.6 & 231.02 & \bf1503.0 & 1541.9 & 105.69 \\
\texttt{RC105\_21x} & \bf1505.0 & 1542.2 & 212.09 & 1511.0 & 1591.9 & 210.93 \\
\texttt{RC106\_21x} & 1512.0 & 1574.9 & 321.03 & \bf1510.0 & 1580.3 & 314.99 \\
\texttt{RC107\_21x} & 1511.0 & 1570.9 & 451.65 & \bf1505.0 & 1568.4 & 211.2 \\
\texttt{RC108\_21x} & \bf1505.0 & 1525.2 & 304.93 & 1511.0 & 1548.9 & 253.96 \\
\texttt{RC201\_21x} & 844.0 & 919.8 & 443.86 & \bf837.0 & 903.9 & 317.43 \\
\texttt{RC202\_21x} & \bf835.0 & 868.6 & 496.73 & \bf835.0 & 881.9 & 219.82 \\
\texttt{RC203\_21x} & 844.0 & 904.0 & 139.53 & \bf841.0 & 908.3 & 274.9 \\
\texttt{RC204\_21x} & 844.0 & 981.6 & 628.28 & \bf840.0 & 891.1 & 341.34 \\
\texttt{RC205\_21x} & 936.0 & 964.3 & 363.54 & \bf813.0 & 889.2 & 389.94 \\
\texttt{RC206\_21x} & 844.0 & 883.9 & 494.53 & \bf841.0 & 878.4 & 328.02 \\
\texttt{RC207\_21x} & 844.0 & 892.5 & 333.73 & \bf829.0 & 897.8 & 335.21 \\
\texttt{RC208\_21x} & 851.0 & 886.0 & 497.78 & \bf838.0 & 898.0 & 600.69 \\
\hline \hline
\texttt{average} & 1154.57 & 1199.07 & 293.88 & \bf1145.5 & 1197.2 & 250.86 \\

\hline
\end{tabular}}
\end{table}

\section{Conclusion and Outlook}
\label{sec:conclusions}

In this study, we presented mathematical formulations of the Two-Echelon Electric Vehicle Routing Problem (2E-EVRP) and various extensions, incorporating real-world constraints such as time windows, satellite synchronization, and pickup and delivery constraints, including simultaneous pickup and deliveries as well as partial deliveries. These extensions were designed to reflect practical logistics challenges, ensuring the applicability of the models in diverse urban logistics scenarios.

We employed a multi-faceted solution strategy to address the complexity of these formulations. The proposed MILP formulations were solved using the commercial solver CPLEX, which provided optimal solutions for small and medium-sized problem instances. However, as expected, solving large-scale instances with exact methods proved computationally challenging, particularly when additional constraints were incorporated. Therefore we applied probabilistic versions of Clark and Wright Savings and Insertion Algorithms as well as metaheuristics such as Variable Neighborhood Search and Construct Merge Solve and Adapt~\cite{blum2024construct}.

By integrating these methodologies, we demonstrated the feasibility of solving large-scale instances of the 2E-EVRP and its extensions. The results highlight the importance of combining exact and heuristic approaches to achieve a practical balance between solution quality and computational efficiency.
\\

\textbf{Acknowledgements.} The research presented in this paper was supported by grants TED2021-129319B-I00 and PID2022-136787NB-I00 funded by MCIN/AEI/10.13039/ 501100011033.

\bibliographystyle{unsrtnat}
\bibliography{paper}  %%% Uncomment this line and comment out the ``thebibliography'' section below to use the external .bib file (using bibtex) .





% \end{thebibliography}

\end{document}